\documentclass[10pt]{article}
\usepackage[margin=1in]{geometry}                
\usepackage{graphicx}  
\usepackage{amsmath}   
\usepackage[compress]{cite}
\usepackage{epstopdf}
\usepackage{amssymb}   
\usepackage{bm} 
\usepackage{dcolumn}
\usepackage{color}
\usepackage{mathrsfs}
\usepackage{amsfonts}
\usepackage{varioref}
\usepackage{textcomp}
\usepackage{enumerate}
\usepackage{paralist}
\usepackage{notes2bib}
\usepackage{mathtools} 
\usepackage{hyperref}
\hypersetup{
	bookmarks=false,         
	pdfstartview={FitH},    
	colorlinks=true,       
	linkcolor=blue,          
	citecolor=red,        
	filecolor=blue,      
	urlcolor=magenta          
}
\allowdisplaybreaks
\DeclareFontFamily{OT1}{pzc}{}
\DeclareFontShape{OT1}{pzc}{m}{it}%
{<-> s * [0.900] pzcmi7t}{}
\DeclareMathAlphabet{\mathscr}{OT1}{pzc}%
{m}{it}
\labelformat{section}{Section #1} 
\labelformat{subsection}{Section #1} 
\labelformat{subsubsection}{Section #1}
\labelformat{subsubsubsection}{Section #1}
\labelformat{equation}{Eq.~(#1)} 
\labelformat{figure}{Fig.~#1} 
\labelformat{subfigure}{Fig.~\thefigure#1} 
\labelformat{table}{Tab.~#1} 
\newcommand{\be}{\begin{equation}}
\newcommand{\ee}{\end{equation}}
\newcommand{\bea}{\begin{eqnarray}}
\newcommand{\eea}{\end{eqnarray}}
\newcommand{\nn}{\nonumber}
\newcommand{\nl}{\hspace{2em}&\hspace{-2em}}

\def\EH{Einstein-Hilbert }
\def\LL{Lanczos-Lovelock }
\def\GHY{Gibbons-Hawking-York }
\def\GB{Gauss-Bonnet }

\def\gr{general relativity}

\newcommand{\sqg}{\sqrt{-g}}
\newcommand{\sqh}{\sqrt{|h|}}

\newcommand{\df}{\delta}

\newcommand{\vbar}{\left\vert\vphantom{\int}\right.}

\newcommand{\sur}{~^{(D-1)}}

\newcommand\Mycomb[2][n]{\prescript{#1\mkern-0.5mu}{}C_{#2}}

\def\pdlr#1#2{\left(\frac{\partial#1}{ \partial#2}\right)}

\def\DM{\mathrm{d}}
\labelformat{section}{Section #1} 
\labelformat{subsection}{Section #1} 
\labelformat{subsubsection}{Section #1}
\labelformat{subsubsubsection}{Section #1}
\labelformat{equation}{Eq.~(#1)} 
\labelformat{figure}{Fig.~#1} 
\labelformat{subfigure}{Fig.~\thefigure#1} 
\labelformat{table}{Tab.~#1} 
\labelformat{appendix}{Appendix #1}
\title{A novel derivation of the boundary term for the action in Lanczos-Lovelock gravity}
\author{Sumanta Chakraborty
\footnote{sumantac.physics@gmail.com}~$^{1}$,
Krishnamohan Parattu
\footnote{mailofkrishnamohan@gmail.com}~$^{2}$
and T. Padmanabhan
\footnote{paddy@iucaa.in}~$^{3}$\\
{$^{1}$\small{Department of Theoretical Physics, Indian Association for the Cultivation of Science, Kolkata-700032, India}}\\
{\small and}\\
{$^{2}$\small{Department of Physics, IIT Madras, Chennai - 600 036, India}}\\
{\small{and}}\\
$^{3}$\small{IUCAA, Post Bag 4, Ganeshkhind, Pune University Campus, Pune 411007, India}}
\begin{document}
  
\maketitle
\begin{abstract}
We present a novel derivation of the boundary term for the action in Lanczos-Lovelock gravity, starting from the boundary contribution in the variation of the Lanczos-Lovelock action. The derivation presented here is straightforward, i.e., one starts from the Lanczos-Lovelock action principle and the action \emph{itself} dictates the boundary structure and hence the boundary term one needs to add to the action to make it well-posed. It also gives the full structure of the contribution at the boundary of the complete action, enabling us to read off the degrees of freedom to be fixed at the boundary, their corresponding conjugate momenta and the total derivative contribution on the boundary. We also provide a separate derivation of the Gauss-Bonnet case. 
\end{abstract}
\section{Introduction}

Einstein-Hilbert action has been the preferred action for general relativity for over a century now, ever since it was introduced by Hilbert and Einstein \cite{Einstein:1916cd} (Einstein's paper also credits previous work done by H. A. Lorentz). It is generally covariant, is constructed out of only the metric and its derivatives and furnishes the Einstein's field equations on variation. Einstein's field equations are in conformity with all well-established experiments and observations \cite{Will:2014kxa}. Any other action which gives the same equations of motion will differ from the Einstein-Hilbert action by a total derivative. Out of all such actions, Einstein-Hilbert action seems the simplest, at least in the metric formulation, and hence has been the preferred action in classical general relativity. 

Although action is just a device to obtain the equations of motion as far as classical physics is concerned, it comes into its own in the realm of quantum physics. Since we live in a quantum world, it does make sense to ask what is the right action among different actions that provide the same equations of motion but differ by total derivatives. In fact, it is in the context of path integral formulation of quantum gravity that Gibbons and Hawking proposed to augment the Einstein-Hilbert action by a boundary term (which can also be obtained by integrating a total derivative over the bulk). This term has come to be known as the Gibbons-Hawking-York term \cite{Gibbons:1976ue,York:1972sj} (also see an earlier paper by Gowdy with essentially the same proposal \cite{Gowdy:1970mz}). This term was added so that normal derivatives of the metric need not be fixed on the boundary as is appropriate in the normal path integral formulation. Also, the variational principle then becomes well-posed (For more 
discussion on the need to add boundary terms to the action, see \cite{Hawking:1980gf, Dyer:2008hb} and Chapter 6 in \cite{gravitation}.) 

The \GHY term is defined in such a way as to be applicable only to a non-null surface. A proposal for the boundary term on a null boundary was recently made \cite{Parattu:2015gga}, which was followed by a proposal for a unified boundary term for null and non-null boundaries \cite{Parattu:2016trq}. This work has been followed up and refined \cite{Chakraborty:2016yna,Lehner:2016vdi,Hopfmuller:2016scf,Jubb:2016qzt}. We will not discuss the null case further in this work. However, note that the \GHY term is not the only term one can add to the action, in principle there are infinite such choices \cite{Charap:1982kn}. In particular, as pointed out in an earlier work of York (and revived recently in \cite{Krishnan:2016mcj,Chakraborty:2016yna}) if one fixes the conjugate momentum on a non-null surface, in four dimensions, one need not have to add any boundary term \cite{York1986}. 

There is a very elegant route to arrive at the Einstein-Hilbert action starting from some very general conditions. If we start assuming that our Lagrangian is to be constructed from the metric and the curvature tensor, and then demand that the equations of motion are second order in the derivatives of the metric, we uniquely arrive at a class of Lagrangians known as Lanczos-Lovelock Lagrangians \cite{lanc38,Lovelock:1971yv,Padmanabhan:2013xyr} (see the introduction of \cite{Deruelle:2003ck} for a taste of history). It is commonly stated that the Lanczos-Lovelock theory, obtained by adding the Lanczos-Lovelock Lagrangians with arbitrary coefficients, reduces to Einstein's theory uniquely when we set the number of dimensions $D=4$. Although this is true at the equations-of-motion level, there are actually two terms that survive at the action level. In addition to the Einstein-Hilbert term, there is also the Gauss-Bonnet term, which is the Lanczos-Lovelock term that is quadratic in the curvature. It is a total 
derivative in four dimensions and hence does not contribute to 
the equations of motion (see Section 2.5 in \cite{Padmanabhan:2013xyr}). This is the reason this term is usually ignored in four dimensions. But when we are concerned with the proper form of the action in four dimensions, this term  
\textit{must} be considered. Of course, it could be that there are more dimensions to the universe than four. If this is the case, the other Lanczos-Lovelock terms also have to be taken into consideration, with the assumption that these higher derivative terms give sub-dominant contribution to the equations of motion which have not been detected in our current experiments and observations. Moreover, there is a claim in the literature that the Gauss-Bonnet term appears as the curvature squared term in the low energy limit of string theory \cite{Zwiebach:1985uq} (also see \cite{Zumino:1985dp}) and hence perhaps other Lovelock terms appear at higher orders (although this was not borne out in the third order calculations done in \cite{Metsaev:1986yb}).

The question of a well-posed variational problem for Lanczos-Lovelock theories has also been tackled in the literature. The appropriate boundary term for Gauss-Bonnet was derived by Bunch \cite{bunch1981surface} (In fact, Bunch demonstrates that only the Gauss-Bonnet Lagrangian among all the Lagrangians constructed out of the quadratic curvature terms $R^2$, $R^{ab}R_{ab}$ and $R^{abcd}R_{abcd}$ has a well-posed boundary value problem.). The boundary terms for general Lanczos-Lovelock was derived by Myers \cite{Myers:1987yn}. This term is generally either derived from topological considerations (looking at the Euler density for a manifold with boundary) or it is demonstrated that it cancels all variations of normal derivatives of the metric on the boundary. In this paper, we attempt to derive these results directly by starting with the boundary terms that appear when we vary the Lanczos-Lovelock Lagrangians. We follow the procedure that was used in \cite{Padmanabhan:2014BT} to derive the \GHY term and then 
in \cite{Parattu:2015gga,Parattu:2016trq} to derive the boundary terms for a null boundary and for a general boundary. After warming up with the Gauss-Bonnet case, we do the calculations in full gory detail for a general \LL theory to separate the boundary variation into the term that has to be canceled by the addition of a boundary term, the terms to be killed by fixing the intrinsic metric on the boundary and the total derivative term on the surface. 

We would like to emphasize that unlike previous attempts, where the boundary term was obtained either by inspection or topological considerations and then shown to cancel the normal derivatives of the metric in the boundary variation, we will follow a more direct root. Starting from the action principle itself and then judiciously manipulating the boundary variations we arrived at the structure of variation on the boundary and hence the boundary term. Thus action principle itself dictates what boundary term one has to add to the action to make it well posed, as well as what one needs to fix on the boundary (for earlier works in the similar spirit, see \cite{Padmanabhan:2014BT,Parattu:2015gga,Parattu:2016trq,Chakraborty:2016yna}). Schematically, the structure one expects due to variation of the action can be presented as
\begin{align}
\delta \left(\int d^{D}x~~\textrm{Lagrangian}\right)=&\int d^{D}x~\left(\textrm{Equation of Motion Term}\right)\delta \left(\textrm{Dynamical Variable} \right)
\nonumber
\\
&+\int d^{D-1}x~~\left(\textrm{Conjugate Momentum}\right)\delta \left(\textrm{Variables to be fixed}\right)
\nonumber
\\
&+\int d^{D-1}x~~\delta \left(\textrm{Boundary Term}\right)+\int d^{D-1}x~~\left(\textrm{Total Divergence Term}\right),
\end{align}
where the first term on the right hand side corresponds to the equations of motion for the dynamical variable. It is clear that in addition to the equation of motion term, one has three additional contributions, the conjugate momentum to the dynamical variable, the boundary term and finally a total derivative term. If one constructs a new action, which is obtained by subtracting out the boundary term from the original action, the variational problem will be well-posed. This will be our aim in this work, i.e., to express the \LL action in the above form, so that the boundary term can be singled out. 

This paper is organized as follows: We give a brief overview of \LL theories in \ref{sec:LL_intro}. The main part of the paper is \ref{sec:BT_LL} where we have derived our boundary term results. In \ref{sec:BT_LL-1}, we manipulate the boundary variation for general \LL stopping just before the point where expressions start to become intense as the determinant tensor makes its entrance. Then, we show in \ref{sec:EH_BT} that the expressions we have derived reproduce the known expressions in the Einstein-Hilbert case. Next, we derive the Gauss-Bonnet case in \ref{sec:Gauss-Bonnet} to obtain a boundary term matching with the one derived by Bunch in \cite{bunch1981surface}. The conjugate momentum is also shown to match with the result previously obtained by Davis \cite{Davis:2002gn} and also by Gravanis and Willison \cite{Gravanis:2002wy}. (The explicit expression is written down in \cite{Deruelle:2003ck}.) \textit{The total derivative term in the boundary variation, as far as we know, has not been written down 
in the literature before.} Finally, in \ref{sec:BT_LL-2}, we complete the full calculation of the decomposition of the general \LL boundary variation. \textit{The total derivative term and the Dirichlet variation term are probably written down in the literature for the first time.} \ref{sec:consistency} is devoted to consistency checks, and we show that the \LL boundary term 
derived 
matches with previous literature and and also that the \LL expressions reduce to the corresponding \EH and \GB expressions for $m=1$ and $m=2$.

The conventions used in this paper are as follows: We use the metric signature $(-,+,+,+)$. The fundamental constants $G$, $\hbar$ and $c$ have been set to unity. The Latin indices, $a,b,\ldots$, run over all space-time indices, and are hence summed over four values (or $D$ values when spacetime is $D$ dimensional). Greek indices, $\alpha ,\beta ,\ldots$, are used when we specialize to indices corresponding to a codimension-1 surface, and are summed over $D-1$ values in $D$ dimensions. $\mathcal{R}$ is used for the the curvature tensor on the boundary surface, while the bulk curvature is represented by $R$. For the connection, $\gamma$ is used for the boundary connection while $\Gamma$, as usual, is used for the bulk connection. The conventions for normal, induced metric, etc. are taken from Appendix B of the arxiv version of \cite{Parattu:2015gga}. The \LL conventions are specified in \ref{sec:LL_intro}.
\section{\LL Theories} \label{sec:LL_intro}

In this section we will rapidly glance through basic aspects of \LL gravity, which will be useful for the later parts of this work. However the interested reader may consult the review \cite{Padmanabhan:2013xyr} for a better understanding of these results. The general \LL action in a volume $\mathcal{V}$ of $D$-dimensional spacetime is given by
\begin{equation} \label{Lovelock_Action}
16\pi\mathcal{A}=\int_{\mathcal{V}} d^{D}x~\sqrt{-g} L_{LL}=\int_{\mathcal{V}} d^{D}x~\sqrt{-g}\sum_{m=1}^{m_{\rm max}}c_m L_m~;\quad L_m\equiv\frac{1}{2^{m}}\delta ^{a_{1}b_{1}\cdots a_{m}b_{m}}_{c_{1}d_{1}\cdots c_{m}d_{m}}R^{c_{1}d_{1}}_{a_{1}b_{1}}\cdots R^{c_{m}d_{m}}_{a_{m}b_{m}}~.
\end{equation}
Here $m_{\rm max}$ is the greatest integer less than or equal to $D/2$. The cosmological constant term has been omitted above, but it may be included as the $m=0$ term. The $16\pi$ and $1/2^m$ factors have been kept separate from the constants $c_m$ so that the $m=1$ term reduces to the standard form of the Einstein-Hilbert action (see Chapter 6 in \cite{gravitation}) when $c_1=1$. 
Here, $\delta ^{a_{1}b_{1}\cdots a_{m}b_{m}}_{c_{1}d_{1}\cdots c_{m}d_{m}}$ is the completely antisymmetric determinant tensor (or alternating tensor) which is defined as the determinant of a matrix made of delta functions as follows:
\begin{eqnarray}
\delta^{i a_1 b_1 \ldots a_m b_m}_{j c_1 d_1 \ldots c_m d_m} = {\mathrm{det}} \left[ \begin{array}{c|ccc}
\delta^i_j & \delta^i_{c_1} & \cdots & \delta^i_{d_m} 
\\
\hline
\\
\delta^{a_1}_j &  & & 
\\
\vdots & & \delta^{a_1 b_1 \ldots a_m b_m}_{c_1 d_1 \ldots c_m d_m} &
\\
\delta^{b_m}_j & & &  \vphantom{\bigg{|}} \end{array} 
\right] \,~.\label{determinant_tensor}
\end{eqnarray}
While working with \LL theories, it is useful to define the tensor $P^{ab}_{cd}$ as
\begin{equation}
P^{ab}_{cd} \equiv \pdlr{L_{LL}}{R_{ab}^{cd}}_{g_{ij}}=\sum _{m}c_{m}\frac{m}{2^{m}}\delta ^{aba_{1}b_{1}\cdots a_{m-1}b_{m-1}}_{cdc_{1}d_{1}\cdots c_{m-1}d_{m-1}}R^{c_{1}d_{1}}_{a_{1}b_{1}}\cdots R^{c_{m-1}d_{m-1}}_{a_{m-1}b_{m-1}}~. \label{P_def}
\end{equation}
which inherits the symmetries from $R_{abcd}$:
\begin{equation}\label{P_symm}
P^{abcd}=-P^{bacd}; \quad P^{abcd}=-P^{abdc};\quad P^{abcd}=P^{cdab}~.
\end{equation}
and have zero divergence, i.e., $\nabla_{a}P^{abcd}=0$. We shall also define the corresponding tensor for $m$th order \LL Lagrangian $L_m$ as
\begin{equation}
P^{ab}_{cd\ (m)} \equiv \pdlr{L_{m}}{R_{ab}^{cd}}_{g_{ij}}~,
\end{equation}
so that for the full \LL Lagrangian one ends up with
\begin{equation}
P^{ab}_{cd}=\sum_m c_m P^{ab}_{cd\ (m)}~.
\end{equation}
The Lagrangian $L_m$ in terms of $P^{ab}_{cd\ (m)}$ becomes
\begin{equation}
L_m =\frac{1}{m}P^{ab}_{cd\ (m)} R^{cd}_{ab}~. 
\end{equation}
For \gr\ ($m=1$), the tensor $P^{ab}_{cd\ (1)}$ becomes
\begin{equation}\label{P_EH}
P^{ab}_{cd\ (1)}=\frac{1}{2}\left(\delta ^{a}_{c}\delta ^{b}_{d}-\delta ^{a}_{d}\delta ^{b}_{c}\right),
\end{equation}
and the Lagrangian is then
\begin{equation} \label{L_EH}
L_1 =P^{ab}_{cd\ (1)} R^{cd}_{ab}=R~. 
\end{equation}
For \GB gravity ($m=2$), we have
\begin{equation}\label{P_Gauss_Bonnet}
P^{ab}_{cd\ (2)}=2\Big[R^{ab}_{cd}+G^{b}_{c}\delta ^{a}_{d}-G^{a}_{c}\df^{b}_{d}+R^{a}_{d}\df^{b}_{c}-R^{b}_{d}\delta ^{a}_{c}\Big]~,
\end{equation}
and
\begin{equation}\label{L_Gauss_Bonnet}
L_2 = \frac{1}{2} P^{ab}_{cd\ (2)} R^{cd}_{ab} = R^2-4 R^{ab}R_{ab}+R^{abcd}R_{abcd}~.
\end{equation}
Keeping these basic results and notations in mind we will now straightforwardly jump into the details of the calculation, where the symmetry properties of the tensor $P^{abcd}$ will be extensively used.

\section{Boundary Variation for the Lanczos-Lovelock Theory}\label{sec:BT_LL}

\subsection{Manipulating the Boundary Term for general Lanczos-Lovelock: Part 1} \label{sec:BT_LL-1}

When the \LL action in \ref{Lovelock_Action} is varied, the boundary term in the variation on a non-null boundary, denoted by $\partial \mathcal{V}$, is given by
\begin{align}
16\pi\mathcal{A}_{\partial\mathcal{V}}&=\int_{\partial\mathcal{V}} d^{D-1}x~\df L_{\partial \mathcal{V}}~;\quad\df L_{\partial \mathcal{V}}=\sqrt{|h|}\mathcal{B}[n_{c}];\label{dL_def} \\
\mathcal{B}[n_{c}]&=2 n_{c}P_{a}^{~bcd}\delta \Gamma ^{a}_{bd}~,\label{B_def}
\end{align}
where the integration is over the boundary, $n_{c}$ is the unit normal to the boundary and $h$ is the determinant of the induced metric $h_{\alpha \beta}$ on the boundary. The conventions used here can be found in Appendix B of the arxiv version of \cite{Parattu:2015gga}. $\mathcal{B}[n_{c}]$ can also be written as
\begin{equation}
\mathcal{B}[n_{c}]= n_{c}P^{mbcd}\left(-\nabla _{m}\delta g_{bd}+\nabla _{b}\delta g_{md}+\nabla _{d}\delta g_{mb}\right)
=2 P^{abcd}n_{c}\nabla _{b}\delta g_{ad}\label{B_def_2}~.
\end{equation}
Introducing the induced metric $$h^{a}_{b}=\delta ^{a}_{b}-\epsilon n^{a}n
_{b},$$ we project out various components of $P^{abcd}$ in \ref{B_def} as follows:
\begin{align}
\mathcal{B}[n_{c}]&=2 n_{c}\delta ^{a}_{m}P_{a}^{~bcd}\delta \Gamma ^{m}_{bd}=2 n_{c}\left(h^{a}_{m}+\epsilon n^{a}n_{m}\right)P_{a}^{~bcd}\delta \Gamma ^{m}_{bd}
\nonumber
\\
&=2 n_{c}h^{a}_{m}\left(h^{b}_{n}+\epsilon n^{b}n_{n}\right)P_{a}^{~ncd}\delta \Gamma ^{m}_{bd} +2\epsilon n_{c}n^{a}n_{m}\left(h^{b}_{n}+\epsilon n^{b}n_{n}\right)P_{a}^{~ncd}\delta \Gamma ^{m}_{bd}
\nonumber
\\
&=2 n_{c}h^{a}_{m}h^{b}_{n}P_{a}^{~ncd}\delta \Gamma ^{m}_{bd}+2\epsilon n_{c}h^{a}_{m}n^{b}n_{n}P_{a}^{~ncd}\delta \Gamma ^{m}_{bd} +2 \epsilon n_{c}n^{a}n_{m}h^{b}_{n}P_{a}^{~ncd}\delta \Gamma ^{m}_{bd}
\nonumber
\\
&=2 n_{c}h^{a}_{m}h^{b}_{n}h^{d}_{p}P_{a}^{~ncp}\delta \Gamma ^{m}_{bd}+2\epsilon n_{c}h^{a}_{m}n^{b}n_{n}h^{d}_{p}P_{a}^{~ncp}\delta \Gamma ^{m}_{bd}+2 \epsilon n_{c}n^{a}n_{m}h^{b}_{n}h^{d}_{p}P_{a}^{~ncp}\delta \Gamma ^{m}_{bd}~. \label{BT_1}
\end{align}
Out of all the possible projections of all indices of $P_{a}^{~bcd}$ using $n_e$ and $h^m_n$, the non-zero ones are the ones with zero, one or two indices contracted with the normal. Contraction of all indices with $h^m_n$ does not contribute in the above expression due to the symmetry in $b$ and $d$.  The contractions that do occur in the above expression are captured by the following surface tensors (i.e. their contraction with $n_a$ on any index is zero): 
\begin{align}
A^{dm}&\equiv 2P^{ancp}n_{c}n_{n}h^{d}_{a}h^{m}_{p}~,\label{def_A} \\
B^{med}&\equiv 2P^{ancp}n_{c}h_{a}^{m}h^e_{n}h^{d}_{p} \label{def_B}
\end{align}
Note that, due to the symmetries of $P^{ancp}$, these are the only contractions possible with one and two contractions along the normal.
We shall also define the corresponding quantities for $P^{ab}_{cd\ (m)}$ as
\begin{align}
A^{dm}_{\ (m)}&\equiv 2P^{ancp}_{\ (m)}n_{c}n_{n}h^{d}_{a}h^{m}_{p}~,\label{def_A_m} \\
B^{med}_{\ (m)}&\equiv 2P^{ancp}_{\ (m)}n_{c}h_{a}^{m}h^e_{n}h^{d}_{p} \label{def_B_m}
\end{align}
so that
\begin{align}
A^{dm}&=\sum_m c_m A^{dm}_{\ (m)}\label{sum_A}~, \\
B^{med}&=\sum_m c_m A^{dm}_{\ (m)} \label{sum_B}~.
\end{align}
The tensor $A^{dm}$ is symmetric: 
\begin{equation}
A^{md}=2P^{ancp}n_{c}n_{n}h^{m}_{a}h^{d}_{p}
=-2P^{cpna}n_{c}n_{n}h^{m}_{a}h^{d}_{p}=2P^{pcna}n_{c}n_{n}h^{m}_{a}h^{d}_{p}=A^{dm}~,
\end{equation}
while the tensor $B^{med}$ is obviously antisymmetric in the first two indices.  
The boundary term in \ref{BT_1} thus becomes
\begin{align}
\mathcal{B}[n_{c}]&=B_{m}^{\phantom{a}bd}\delta \Gamma ^{m}_{bd} +\epsilon \left(A^{d}_{m}n^{b}\delta \Gamma ^{m}_{db}-A^{bd}n_{m}\delta \Gamma ^{m}_{bd}\right)
\nonumber
\\
&=\mathcal{B}_{1}[n_{c}]+\mathcal{B}_{2}[n_{c}]~. \label{BT_2}
\end{align}
The first term $\mathcal{B}_{1}[n_{c}]$ can be rewritten in terms of the covariant derivative of variations of the metric as
\begin{align}
\mathcal{B}_{1}[n_{c}]&=B_{m}^{\phantom{a}bd}\left[\frac{g^{mq}}{2}\left(-\nabla _{q}\delta g_{bd}+\nabla _{b}\delta g_{dq}+\nabla _{d}\delta g_{bq}\right)\right]=B^{qbd}\nabla _{b}\delta g_{dq}
\nonumber
\\
&=\nabla _{b}\left\lbrace B^{qbd}\delta g_{dq} \right\rbrace
-(\nabla _{b} B^{qbd})\delta g_{dq}
=\nabla _{b}\left\lbrace B^{qbd}\delta h_{dq} \right\rbrace
+(\nabla^{b} B_{qbd})\delta g^{dq}
\nonumber
\\
&=\nabla _{b}\left\lbrace B^{qbd}\delta h_{dq} \right\rbrace
+(\nabla^{b} B_{qbd})\delta h^{dq} 
+\epsilon\left(n^{d}\delta n^{q}+n^{q}\delta n^{d} \right)\nabla^{b} B_{qbd}~. \label{B1}
\end{align}
In the second line above, we have used the fact that the contraction of $B^{qbd}$ with $n_a$ on any index is zero to convert $\df g_{dq}$ to $\df h_{dq}$. The last two terms can be simplified using the following results. We have
\begin{align}
n^{d}\nabla^{b} B_{qbd}&=n^{d}\nabla _{b}\left\lbrace 2 h^{b}_{n}P^{ancp}h_{aq}n_{c}h_{pd}\right\rbrace
=-2 h^{b}_{n}P^{ancp}h_{aq}n_{c}h_{pd}\nabla _{b}n^{d}\nn\\
&=2h^{b}_{n}P^{ancp}h_{aq}n_{c}h_{pd}K^{d}_{b}=B_{qbd}K^{bd}~,
\end{align}
where we have used the relation
\begin{equation}\label{dn_K}
\nabla _{a}n^{b}=-K^{b}_{a}+\epsilon n_{a}a^{b}~,
\end{equation}
with the acceleration $a^b=n^a\nabla_a n^b$. The next result is
\begin{align}
n^{q}\nabla^{b} B_{qbd}=n^{q}\nabla _{b}\left\lbrace 2 h^{b}_{n}P^{ancp}h_{aq}n_{c}h_{pd}\right\rbrace&=-2 h^{b}_{n}P^{ancp}h_{aq}n_{c}h_{pd}\nabla _{b}n^{q}=2P^{ancp}K_{an}n_{c}h_{pd}=0~,
\end{align}
using \ref{P_symm}. Thus, we obtain
\begin{equation}
\mathcal{B}_{1}[n_{c}]=\nabla _{b}\left\lbrace B^{qbd}\delta h_{dq} \right\rbrace
+(\nabla^{b} B_{qbd})\delta h^{dq} 
+\epsilon B_{qbd}K^{bd} \delta n^{q}~.
\end{equation}
From the first term, we can separate out the surface covariant derivative.
\begin{align}
\nabla _{b}\left\lbrace B^{qbd}\delta h_{dq} \right\rbrace &= h^c_b \nabla _{c}\left\lbrace B^{qbd}\delta h_{dq} \right\rbrace+\epsilon n^c n_b \nabla _{c}\left\lbrace B^{qbd}\delta h_{dq} \right\rbrace~ \nn \\
&=h^c_b \nabla _{c}\left\lbrace B^{qbd}\delta h_{dq} \right\rbrace-\epsilon n^c  B^{qbd}\left\lbrace\nabla _{c}n_b\right\rbrace\delta h_{dq}\nn \\
&= D_{b}\left\lbrace B^{qbd}\delta h_{dq} \right\rbrace-\epsilon n^c  B^{qbd}\left\lbrace-K_{cb}+\epsilon n_{c}a_{b}\right\rbrace\delta h_{dq} \nn\\
&=D_{b}\left\lbrace B^{qbd}\delta h_{dq} \right\rbrace-\epsilon B^{qbd}a_{b}\delta h_{dq},
\end{align}
where we have used \ref{dn_K}. 
Thus, $\mathcal{B}_{1}[n_{c}]$ becomes
\begin{align}
\mathcal{B}_{1}[n_{c}]&=D_{b}\left\lbrace B^{qbd}\delta h_{dq} \right\rbrace-\epsilon B^{qbd}a_{b}\delta h_{dq}
+(\nabla^{b} B_{qbd})\delta h^{dq} 
+ \epsilon B_{qbd}K^{bd}\delta n^{q}~ \nn \\
&=D_{b}\left\lbrace B^{qbd}\delta h_{dq} \right\rbrace-\epsilon B^{qbd}a_{b}\delta g_{dq}
+(\nabla^{b} B_{qbd})\delta h^{dq} 
+ \epsilon B_{qbd}K^{bd}\delta n^{q}~ \nn \\
&=D_{b}\left\lbrace B^{qbd}\delta h_{dq} \right\rbrace+\epsilon B_{qbd}a^{b}\delta g^{dq}
+(\nabla^{b} B_{qbd})\delta h^{dq} 
+ \epsilon B_{qbd}K^{bd}\delta n^{q}~ \nn \\
&=D_{b}\left\lbrace B^{qbd}\delta h_{dq} \right\rbrace+\epsilon B_{qbd}a^{b}\delta h^{dq}
+(\nabla^{b} B_{qbd})\delta h^{dq} 
+ \epsilon B_{qbd}K^{bd}\delta n^{q}~ \nn \\
&=D_{b}\left\lbrace B^{qbd}\delta h_{dq} \right\rbrace+\left(
\nabla^{b} B_{qbd}+\epsilon B_{qbd}a^{b}\right)\delta h^{dq} 
+ \epsilon B_{qbd}K^{bd}\delta n^{q}~ \nn \\
&=D_{b}\left\lbrace B^{qbd}\delta h_{dq} \right\rbrace+\left(
D^{b} B_{qbd}\right)\delta h^{dq} 
+ \epsilon B_{qbd}K^{bd}\delta n^{q}~.
\label{B1_simplified_1}
\end{align}
In the last line above, we have used the result
\begin{align}
D^{b} B_{qbd}=h^{bc}\nabla_c B_{qbd}
=\nabla^b B_{qbd}-\epsilon n^b n^c\nabla_c B_{qbd}
=\nabla^b B_{qbd}+\epsilon  B_{qbd} n^c\nabla_c n^b
=\nabla^b B_{qbd}+\epsilon  B_{qbd} a^b~. 
\end{align}
We shall now turn our attention to the second term, $\mathcal{B}_{2}[n_{c}]$, in \ref{BT_2}: 
\begin{align}
\mathcal{B}_{2}[n_{c}]&=\epsilon \left(A^{d}_{m}n^{b}\delta \Gamma ^{m}_{db}-A^{bd}n_{m}\delta \Gamma ^{m}_{bd}\right)
\nonumber
\\
&=\epsilon\left\{A^{d}_{m}\left[ \delta \left(\nabla _{d}n^{m}\right)-\nabla _{d}\delta n^{m} \right]
-A^{bd}\left[\nabla _{b}\delta n_{d}-\delta \left(\nabla _{b}n_{d}\right) \right]\right\}
\nonumber
\\
&=\epsilon\left\{A^{d}_{m}\delta \left(\nabla _{d}n^{m}\right)+A^{bd}\delta \left(g_{dq}\nabla _{b}n^{q}\right)
-A^{d}_{m}\nabla _{d}\left(\delta n^{m}+g^{mq}\delta n_{q}\right)\right\}
\nonumber
\\
&=\epsilon\left\{2A^{d}_{m}\delta \left(\nabla _{d}n^{m}\right)+A^{bd}\left(\nabla _{b}n^{q}\right)\delta g_{dq}-A^{d}_{m}\nabla _{d}\delta u^{m}\right\},
\end{align}
where we have introduced $\delta u^{m}=\delta n^{m}+g^{mq}\delta n_{q}$, which is a surface vector, i.e. $\df u^m n_m=0$. Using \ref{dn_K}, we obtain
\begin{align}
\mathcal{B}_{2}[n_{c}]&=\epsilon \left\{2A^{d}_{m}\delta \left(-K^{m}_{d}+\epsilon n_{d}a^{m}\right)-A^{b}_{d}\left(\nabla _{b}n_{q}\right)\delta \left(h^{dq}+\epsilon n^{d}n^{q}\right)-A^{d}_{m}\nabla _{d}\delta u^{m}\right\}
\nonumber
\\
&=\epsilon \left\{-2A^{d}_{m}\delta K^{m}_{d}-A^{b}_{d}\left(\nabla _{b}n_{q}\right)\delta h^{dq}-A^{d}_{m}\nabla _{d}\delta u^{m}\right\},
\end{align}
where we have used $\delta n_{d}\propto n_{d}$, $A^{d}_{m}n_{d}=0$ and $n^{c}\nabla _{b}n_{c}=(1/2)\nabla _{b}n^{2}=0$. We can separate out a surface covariant derivative from the last term as follows:
\begin{align}
-A^{d}_{m}\nabla _{d}\delta u^{m}&=-A^{p}_{q}\left(h^{d}_{p}+\epsilon n^{d}n_{p}\right)\left(h^{q}_{m}+\epsilon n^{q}n_{m}\right)\nabla _{d}\delta u^{m}
\nonumber
\\
&=-A^{p}_{q}D_{p}\delta u^{q}=-D_{p}\left(A^{p}_{q}\delta u^{q}\right)+\delta u^{q}\left(D_{p}A^{p}_{q}\right)~. \label{24}
\end{align}
Evaluating $D_{p}A^{p}_{q}$ using the definition of $A^{p}_{q}$ from \ref{def_A}, we have
\begin{align}
D_{p}A^{p}_{q}&=D_{p}\left(2P_{a}^{~bcd}n_{c}n_{b}h^{a}_{q}h^{p}_{d} \right)
=h^{e}_{p}h^{m}_{q}\nabla _{e}\left(2P_{a}^{~bcd}n_{c}n_{b}h^{a}_{m}h^{p}_{d}\right)=-2P_{a}^{~bcd}n_{b}K_{dc}h^{a}_{q}
-2P_{a}^{~bcd}n_{c}K_{db}h^{a}_{q}
\nonumber
\\
&=-2P_{a}^{~bcd}n_{c}K_{db}h^{a}_{q}
=-2P^{abcd}h_{aq}n_{c}h_{d}^{p}h_{b}^{n}K_{pn}=- B_{qpn}K^{pn}~.
\end{align}
Therefore, the last term in \ref{24} will become
\begin{equation}
\delta u^{q}\left(D_{p}A^{p}_{q}\right)=- B_{qpn}K^{pn}\delta u^{q}=- B_{qpn}K^{pn}\left(\delta n^{q}+g^{qm}\df n_m\right)=- B_{qpn}K^{pn}\delta n^{q}~,
\end{equation}
where we have used the fact that $\df n_m\propto n_m$. Thus, we arrive at the following expression for $\mathcal{B}_{2}[n_{c}]$:
\begin{equation}
\mathcal{B}_{2}[n_{c}]=\epsilon \left\{-2A^{d}_{m}\delta K^{m}_{d}-A^{b}_{d}\left(\nabla _{b}n_{q}\right)\df h^{dq}-D_{p}\left(A^{p}_{q}\delta u^{q}\right)- B_{qpn}K^{pn}\delta n^{q}\right\}~.\label{B2_final}
\end{equation}
Adding \ref{B1_simplified_1} and \ref{B2_final}, the boundary term for \LL gravity, sans the $\sqh$ factor, can be written in the form
\begin{align}
\mathcal{B}[n_{c}]&=D_{b}\left( B^{qbd}\delta h_{dq}-\epsilon A^{b}_{q}\delta u^{q} \right)-2\epsilon A^{d}_{m}\delta K^{m}_{d}+\left(
D^{b} B_{qbd}-\epsilon A^{b}_{d}\nabla _{b}n_{q}\right)\delta h^{dq} \nn \\
&=D_{b}\left( B^{qbd}\delta h_{dq}-\epsilon A^{b}_{q}\delta u^{q} \right)-2\epsilon A^{d}_{m}\delta K^{m}_{d}+\left(
D^{b} B_{qbd}+\epsilon A^{b}_{d}K _{bq}\right)\delta h^{dq} \label{BT_LL_1}~,
\end{align}
where we have used \ref{dn_K} on the last term. The first and last terms are of desired form. Once we add the $\sqh$ factor, the first will give a total derivative on the boundary surface. The last term can be killed by fixing the intrinsic metric on the boundary. The term $-2\epsilon A^{d}_{m}\delta K^{m}_{d}$ has to be manipulated further. Using \ref{sum_A} and \ref{sum_B}, the above boundary variation can be written as
\begin{align}
\mathcal{B}[n_{c}]= \sum_m c_m \left[D_{b}\left( B^{qbd}_{\ (m)}\delta h_{dq}-\epsilon A^{b}_{q\ (m)}\delta u^{q} \right)-2\epsilon A^{d}_{m\ (m)}\delta K^{m}_{d}+\left(
D^{b} B_{qbd\ (m)}+\epsilon A^{b}_{d\ (m)}K _{bq}\right)\delta h^{dq}\right] \label{BT_LL_1_sum}~,
\end{align}
We shall now reproduce the known result in the Einstein-Hilbert case ($m=1$ in the above sum) before deriving the complete boundary variation for the Gauss-Bonnet case ($m=2$). The boundary term for general \LL is derived in \ref{sec:BT_LL-2}.
\subsection{Einstein-Hilbert Case}\label{sec:EH_BT}

In this section, we shall evaluate the boundary variation derived in \ref{BT_LL_1} for the Einstein-Hilbert case and compare it with the results previously known (re derived in \cite{Padmanabhan:2014BT}, Appendix B in the arxiv version of \cite{Parattu:2015gga}). For Einstein-Hilbert, using \ref{P_EH} in \ref{def_A_m} and \ref{def_B_m}, we obtain
\begin{equation}
A^{ef}_{(1)}=2P^{abcd}_{(1)}n_{c}n_{b}h^{e}_{a}h^{f}_{d}=\left(g ^{ac}g ^{bd}-g^{ad}g ^{bc}\right)n_{c}n_{b}h^{e}_{a}h^{f}_{d}=-\epsilon h^{ef}~,
\end{equation}
and
\begin{equation}
B^{efg}_{(1)}=2P^{abcd}_{(1)}n_{c}h^e_{a}h^{f}_{b}h^{g}_{d}=0~.
\end{equation}
Thus, \ref{BT_LL_1} becomes
\begin{align}
\mathcal{B}[n_{c}]&=D_{b}\left( h^{b}_{q}\delta u^{q} \right)+2 h^{d}_{m}\delta K^{m}_{d}-
 h^{b}_{d}K_{bq}\delta h^{dq} \nn\\
 &= D_{b}\left(\delta u^{b}\right)+2 \delta K -K_{dq}\delta h^{dq}~.
 \label{BT_EH}
\end{align}
Adding a $\sqh$ and manipulating, we will arrive at the full boundary variation for the Einstein-Hilbert action as integral over the boundary of
\begin{align}
\sqh\mathcal{B}[n_{c}]= D_{b}\left(\sqh\delta u^{b}\right)+2 \delta \left(\sqh K\right) -\sqh \left(K_{dq}-Kh_{dq}\right)\delta h^{dq}~.
\label{BT_EH_final}
\end{align}
This matches with expressions previously obtained in the literature (see \cite{Padmanabhan:2014BT} as well as Appendix B in the arxiv version of \cite{Parattu:2015gga}). The last term of \ref{BT_EH_final} can be rewritten as $\sqh \left(K^{dq}-Kh^{dq}\right)\delta h_{dq}$, allowing us to identify $K^{dq}-Kh^{dq}$ as the Brown-York stress tensor \cite{Brown:1992br,Bose:1998uu}. The Brown-York quasi-local energy derived from the stress tensor has essentially the same form as the boundary term, with the difference that the integral is that of the extrinsic curvature of a surface of co-dimension 2, over that surface. (The corresponding quantity  embedded in flat spacetime is subtracted out for finiteness, when required.) Recently, this fact was used to extend the Brown-York formulation to Lanczos-Lovelock theories of gravity in \cite{Chakraborty:2015kva} using the boundary terms available in the literature. Thus, obtaining the structure of the boundary variation in Lanczos-Lovelock gravity will also allow us to explore the Brown-York formulation in Lanczos-Lovelock theories of gravity.
\subsection{Gauss-Bonnet Gravity}\label{sec:Gauss-Bonnet}

For \GB gravity, using \ref{P_Gauss_Bonnet} in \ref{def_A_m}, we get
\begin{align}
A^{d}_{m\ (2)}&=-4R_{anpc}h^{a}_{m}n^{n}h^{pd}n^{c}+4G^{nc}n_{c}n_{n}h^{d}_{m}+4\epsilon R^{pa}h_{am}h^{d}_{p}
\nonumber
\\
&=4\epsilon \left[R^{pa}h_{am}h^{d}_{p}-\epsilon R_{anpc}h^{a}_{m}n^{n}h^{pd}n^{c} \right]+4G^{nc}n_{c}n_{n}h^{d}_{m}~.
\nonumber
\end{align}
We shall now use the following results from Chapter 12 in \cite{gravitation} (where the $\epsilon=-1$ case has been derived): 
\begin{align}
R_{ms}h^m_a h^s_c-\epsilon h^m_a h^s_c n^n n^t R_{mnst}&=\mathcal{R}_{ac}-\epsilon\left(K_{ac} K-K_a^b K_{bc}\right); \\
G^{bc}n_b n_c&=\frac{1}{2}\left(K^2-K_{mn}K^{mn}-\epsilon \mathcal{R}\right)~.
\end{align}
Substituting, we obtain
\begin{align}
A^{d}_{m\ (2)}&=4\epsilon \left(\mathcal{R}^{d}_{m}-\frac{1}{2}\mathcal{R}h^{d}_{m}\right)-4\left(KK^{d}_{m}-K^{d}_{c}K^{c}_{m}\right)
+2\left(K^{2}-K_{ab}K^{ab}\right)h^{d}_{m}~. \label{Adm_GB}
\end{align}
Therefore,
\begin{align}
-2A^{d}_{m\ (2)}\delta K^{m}_{d}=&-8\epsilon \left(\mathcal{R}^{d}_{m}-\frac{1}{2}\mathcal{R}h^{d}_{m}\right)\delta K^{m}_{d}
+8\left(KK^{d}_{m}-K^{d}_{c}K^{c}_{m}\right)\delta K^{m}_{d}-4\left(K^{2}-K_{ab}K^{ab}\right)h^{d}_{m}\delta K^{m}_{d}
\nonumber
\\
=&\delta \left(-8\epsilon \mathcal{R}^{d}_{m}K^{m}_{d}+4\epsilon \mathcal{R}K\right)
+8\epsilon K^{m}_{d}\delta \left(\mathcal{R}^{d}_{m}-\frac{1}{2}\mathcal{R}h^{d}_{m}\right)
\nonumber
\\
&-\delta \left(\frac{8}{3}K^{d}_{c}K^{c}_{m}K^{m}_{d}\right)+8KK^{d}_{m}\delta K^{m}_{d}
-4\left(K^{2}-K_{ab}K^{ab}\right)\delta K+4\left(K^{2}-K_{ab}K^{ab}\right)K^{m}_{d}\delta h^{d}_{m}
\nonumber
\\
=&\delta \left(-8\epsilon \mathcal{R}^{d}_{m}K^{m}_{d}+4\epsilon \mathcal{R}K- \frac{8}{3}K^{d}_{c}K^{c}_{m}K^{m}_{d}\right)
+8\epsilon K^{m}_{d}\delta \left(\mathcal{R}^{d}_{m}-\frac{1}{2}\mathcal{R}h^{d}_{m}\right)
\nonumber
\\
&+8KK^{d}_{m}\delta K^{m}_{d}
-\frac{4}{3}\delta \left(K^{3}\right)+4K_{ab}K^{ab}\delta K
\nonumber
\\
=&\delta \left(-8\epsilon \mathcal{R}^{d}_{m}K^{m}_{d}+4\epsilon \mathcal{R}K-\frac{8}{3}K^{d}_{c}K^{c}_{m}K^{m}_{d}-\frac{4}{3}K^{3}+4KK^{a}_{b}K^{b}_{a}\right)
\nonumber
\\
&+8\epsilon K^{m}_{d}\delta \left(\mathcal{R}^{d}_{m}-\frac{1}{2}\mathcal{R}h^{d}_{m}\right)
\nn \\
=&\delta \left(-8\epsilon \mathcal{R}^{d}_{m}K^{m}_{d}+4\epsilon \mathcal{R}K-\frac{8}{3}K^{d}_{c}K^{c}_{m}K^{m}_{d}-\frac{4}{3}K^{3}+4KK^{a}_{b}K^{b}_{a}\right)
\nonumber
\\
&+8\epsilon K^{m}_{d}\delta \left[\mathcal{R}^{d}_{m}\right]-4\epsilon K \delta\left[\mathcal{R}\right]
\nn \\
=&\delta \left(-8\epsilon \mathcal{R}^{d}_{m}K^{m}_{d}+4\epsilon \mathcal{R}K-\frac{8}{3}K^{d}_{c}K^{c}_{m}K^{m}_{d}-\frac{4}{3}K^{3}+4KK^{a}_{b}K^{b}_{a}\right)
\nonumber
\\
&+8\epsilon \left(K^{m}_{d}\mathcal{R}_{mp}-\frac{K}{2}\mathcal{R}_{dp}\right)\delta h^{dp}+8\epsilon \left(K^{ab}-\frac{K}{2}h^{ab}\right)\delta \left[\mathcal{R}_{ab}\right]
\end{align}
Substituting this in \ref{BT_LL_1}, we obtain
\begin{align}
\mathcal{B}[n_{c}]=&D_{b}\left( B^{qbd}_{\ (2)}\delta h_{dq}-\epsilon A^{b}_{q\ (2)}\delta u^{q} \right)
\nonumber
\\
&+\delta \Big[-8 \mathcal{R}^{a}_{b}K^{b}_{a}+4 \mathcal{R}K-\epsilon\left(\frac{4}{3}K^{3}+\frac{8}{3}K^{a}_{b}K^{b}_{c}K^{c}_{a}-4KK^{a}_{b}K^{b}_{a}\right)\Big]
\nonumber
\\
&+\Big[D^{b} B_{qbd\ (2)}+\epsilon A^{b}_{d\ (2)}K _{bq}+8 \left(K^{m}_{d}\mathcal{R}_{mq}-\frac{K}{2}\mathcal{R}_{dq}\right)
\Big]\delta h^{dq}+8 \left(K^{ab}-\frac{K}{2}h^{ab}\right)\delta \left[\mathcal{R}_{ab}\right]~.\label{GB_BT_1}
\end{align}
The second line has the term that will give the term to be canceled by the boundary term (once multiplied by $\sqh$). The first term will give a total derivative on the boundary surface once multiplied by $\sqrt{h}$, while the other terms contain only variations of the boundary metric and will be killed once the induced metric on the boundary is fixed. The $\delta \mathcal{R}_{ab}$
term may be further manipulated to write it as terms containing variations of only the induced metric and those containing the variations of the surface derivatives of the metric.
 
First, we will work with $S^{ab}\delta\mathcal{R}_{ab}$, such that $S^{ab}$ is symmetric and $n_{a}S^{ab}=0$, to derive a general expression that we will later specialize to our case. Further, we shall introduce the convenient notation $\gamma ^{a}_{bc}=\sur \Gamma ^{a}_{bc}$ for the Christoffel symbols of the boundary metric. We use the boundary adapted normal coordinate system (BNC) (see \ref{app:coord}) and write
\begin{align}
S^{ab}\delta\left[\mathcal{R}_{ab}\right]=S^{\alpha \beta}\delta\left[\mathcal{R}_{\alpha \beta}\right]&=S^{\alpha \beta}\left(D_{\mu}\delta \gamma ^{\mu}_{\alpha \beta}-D_{\beta}\delta \gamma ^{\mu}_{\alpha \mu}\right)
\nonumber
\\
&=D_{\mu}\left(S^{\alpha \beta}\delta \gamma ^{\mu}_{\alpha \beta}-S^{\alpha \mu}\delta \gamma ^{\beta}_{\alpha \beta}\right)-\left(D_{\mu}S^{\alpha \beta}\right)\delta \gamma ^{\mu}_{\alpha \beta}
+\left(D_{\beta}S^{\alpha \beta}\right)\delta \gamma^{\mu}_{\alpha \mu}, \label{TR}
\end{align}
where the Greek alphabets indicate indices running over $D-1$ values that correspond to coordinates on the boundary surface. The advantage of doing the calculations in this coordinate system is that all the relations of Riemannian geometry, like the relation for $\delta\left[\mathcal{R}_{\alpha \beta}\right]$ above, can be directly used if we take the viewpoint of a bug which is on the boundary and is not aware of the normal dimension. We shall do our calculations in BNC and then upgrade the results to bulk relations using the procedure in \ref{app:bulk-boundary_correspondence}. Variation of the three-dimensional connection is
\begin{align}
\delta \gamma ^{\mu}_{\alpha \beta}&=\frac{h^{\mu \nu}}{2}\left(-D_{\nu}\delta h_{\alpha \beta}+D_{\alpha}\delta h_{\beta\nu}+D_{\beta}\delta h_{\alpha\nu}\right), \label{3gamma_1}
\\
\delta \gamma ^{\mu}_{\alpha \mu}&=\frac{h^{\mu \nu}}{2}D_{\alpha}\delta h_{\mu \nu}~.\label{3gamma_2}
\end{align}
This can be upgraded to have the bulk relation
\begin{align}
\delta \overline{\gamma} ^{m}_{a b}&\equiv \frac{h^{m n}}{2}\left[-D_{n}\left(h^c_{a}h^d_{b}\delta h_{cd}\right)+D_{a}\left(h^c_{b}h^d_{n}\delta h_{cd}\right) +D_{b}\left(h^c_{a}h^d_{n}\delta h_{cd}\right)\right], \label{3gamma_1_u}
\\
\delta \overline{\gamma} ^{m}_{a m}&= \frac{h^{m n}}{2}D_{a}\left(h^c_{m}h^d_{n}\delta h_{cd}\right)~,\label{3gamma_2_u}
\end{align}
where we have defined the object $\df\overline{\gamma}^{a}_{bc}$. This is not the variation of the object $\gamma^{a}_{bc}=(h^{ad}/2)(-\partial_d h_{bc}+\partial_b h_{cd}+\partial_c h_{bd})$, which is the natural extension of $\gamma^{\alpha}_{\beta \gamma}$ to the bulk. The variation of $\gamma^{a}_{bc}$ is in fact not a tensor (see \ref{app:gamma_gamma_bar}).
Using \ref{3gamma_1} and \ref{3gamma_2}, we can write \ref{TR} as
\begin{align}
-\left(D_{\mu}S^{\alpha \beta}\right)\delta \gamma ^{\mu}_{\alpha \beta}&=\frac{D^{\nu}S^{\alpha \beta}}{2}\left(D_{\nu}\delta h_{\alpha \beta}-D_{\alpha}\delta h_{\beta \nu}-D_{\beta}\delta h_{\alpha \nu}\right)
\nonumber
\\
&=D_{\alpha}\left[ -\left(D^{\nu}S^{\alpha \beta}\right)\delta h_{\beta \nu}+\frac{1}{2}\left(D^{\alpha}S^{\mu \nu}\right)\delta h_{\mu \nu}\right]
+\left[ \frac{1}{2}\left(D_{\mu}D^{\mu}S_{\alpha \beta}\right)-D_{\mu}D_{\alpha}S^{\mu}_{\beta}\right] \delta h^{\alpha \beta},
\end{align}
and
\begin{align}
\left(D_{\beta}S^{\alpha \beta}\right)\delta \gamma ^{\mu}_{\alpha \mu}&=\frac{1}{2}\left(D_{\beta}S^{\alpha \beta}\right)h^{\mu \nu}D_{\alpha}\delta h_{\mu \nu}=D_{\alpha}\left[\frac{1}{2}\left(D_{\beta}S^{\alpha \beta}\right)h^{\mu \nu}\delta h_{\mu \nu} \right] +D_{\mu}\left[ \frac{1}{2} \left(D_{\nu}S^{\mu \nu}\right)h_{\alpha \beta} \right] \delta h^{\alpha \beta}~.
\end{align}
Substituting in \ref{TR}, we have the following relation valid in BNC:
\begin{align}
S^{ab}\delta\left[\mathcal{R}_{ab}\right]=&D_{\mu}\left(S^{\alpha \beta}\delta \gamma ^{\mu}_{\alpha \beta}-S^{\alpha \mu}\delta \gamma ^{\beta}_{\alpha \beta}\right)+D_{\alpha}\left[ -\left(D^{\nu}S^{\alpha \beta}\right)\delta h_{\beta \nu}+\frac{1}{2}\left(D^{\alpha}S^{\mu \nu}\right)\delta h_{\mu \nu}\right]
\nn\\
&+\left[ \frac{D_{\mu}D^{\mu}S_{\alpha \beta}}{2}-D_{\mu}D_{\alpha}S^{\mu}_{\beta}\right] \delta h^{\alpha \beta}+D_{\alpha}\left[\left(\frac{D_{\beta}S^{\alpha \beta}}{2}\right)h^{\mu \nu}\delta h_{\mu \nu} \right] +D_{\mu}\left[\frac{D_{\nu}S^{\mu \nu}h_{\alpha \beta}}{2}\right] \delta h^{\alpha \beta}~.
\end{align}
The corresponding tensorial result in terms of bulk quantities can be written down making use of the prescription in \ref{app:bulk-boundary_correspondence} and $\df\overline{\gamma}^{a}_{bc}$ defined in \ref{3gamma_1_u}:
\begin{align}
S^{ab}\delta\left[\mathcal{R}_{ab}\right]=&D_{m}\left(S^{a b}\delta \overline{\gamma} ^{m}_{a b}-S^{a m}\delta \overline{\gamma} ^{b}_{a b}\right)+D_{a}\left[ -\left(D^{n}S^{a b}\right)\delta h_{bn}+\frac{1}{2}\left(D^{a}S^{m n}\right)\delta h_{mn}\right]
\nn\\
&+\left[ \frac{D_{m}D^{m}S_{a b}}{2}-D_{m}D_{a}S^{m}_{b}\right] \delta h^{a b}+D_{a}\left[\left(\frac{D_{b}S^{a b}}{2}\right)h^{m n}\delta h_{mn} \right]+D_{m}\left[\frac{D_{n}S^{m n}h_{a b}}{2}\right] \delta h^{a b}~.
\end{align}
Substituting these expressions in \ref{TR} and then substituting the resulting expression in \ref{GB_BT_1}, we finally obtain the desired form for $\mathcal{B}[n_{c}]$ in Gauss-Bonnet:  
\begin{align}
\mathcal{B}[n_{c}]=& D_{b}\left\lbrace B^{qbd}_{\ (2)}\delta h_{dq}-\epsilon A^{b}_{q\ (2)}\delta u^{q}+\left(S^{pq}\delta \overline{\gamma} ^{b}_{pq}-S^{ab}\delta \overline{\gamma} ^{c}_{ac}\right)-\left[D^{d}S^{bc}-\frac{1}{2}\left(D^{b}S^{cd}+h^{cd}D_{a}S^{ab}\right) \right]\delta h_{cd}\right\rbrace
\nonumber
\\
&+\delta \Big[-8 \mathcal{R}^{a}_{b}K^{b}_{a}+4 \mathcal{R}K-\epsilon\left(\frac{4}{3}K^{3}+\frac{8}{3}K^{a}_{b}K^{b}_{c}K^{c}_{a}-4KK^{a}_{b}K^{b}_{a}\right)\Big]
\nonumber
\\
&+\Big[D^{b} B_{qbd\ (2)}+\epsilon A^{b}_{d\ (2)}K _{bq}+8 \left(K^{m}_{d}\mathcal{R}_{mq}-\frac{K}{2}\mathcal{R}_{dq}\right)\nn\\
&\phantom{+\Big[}+ \frac{1}{2} \left(h_{dq}D_{a} D_{b}S^{ab}+D_{c}D^{c}S_{dq}\right) - D_{c}D_{d}S^{c}_{q}
\Big]\delta h^{dq}~,\label{GB_BT_final}
\end{align}
where $S^{ab}=8(K^{ab}-(1/2)Kh^{ab})$. Adding the $\sqh$ factor, we have the boundary term to be added to the action as the integral over the boundary of
\begin{equation}\label{GB_boundary_term}
\sqh BT_2 = \sqh \left[8 \mathcal{R}^{a}_{b}K^{b}_{a}-4 \mathcal{R}K+\epsilon\left(\frac{4}{3}K^{3}+\frac{8}{3}K^{a}_{b}K^{b}_{c}K^{c}_{a}-4KK^{a}_{b}K^{b}_{a}\right)\right]~
\end{equation}
and the full boundary variation as
\begin{align}
\nl\sqh\mathcal{B}[n_{c}]\nn\\=& \sqh D_{b}\left\lbrace B^{qbd}_{\ (2)}\delta h_{dq}-\epsilon A^{b}_{q\ (2)}\delta u^{q}+\left(S^{pq}\delta \overline{\gamma} ^{b}_{pq}-S^{ab}\delta \overline{\gamma} ^{c}_{ac}\right)-\left[D^{d}S^{bc}-\frac{1}{2}\left(D^{b}S^{cd}+h^{cd}D_{a}S^{ab}\right) \right]\delta h_{cd}\right\rbrace
\nonumber
\\
&-\delta\left\{\sqh BT_2\right\}
\nonumber
\\
&+\sqh\Big[D^{b} B_{qbd\ (2)}+\epsilon A^{b}_{d\ (2)}K _{bq}+8 \left(K^{m}_{d}\mathcal{R}_{mq}-\frac{K}{2}\mathcal{R}_{dq}\right)\nn\\
&\phantom{+\sqh\Big[]}+ \frac{1}{2} \left(h_{dq}D_{a} D_{b}S^{ab}+D_{c}D^{c}S_{dq}\right) - D_{c}D_{d}S^{c}_{q}-\frac{1}{2}BT_2 h_{dq}
\Big]\delta h^{dq}~.\label{GB_BT_final_sqh}
\end{align}
The boundary term in \ref{GB_boundary_term} matches the one given by Yale \cite{Yale:2011dq} (for a timelike boundary with $\epsilon=+1$) citing Bunch \cite{bunch1981surface}. In the original source \cite{bunch1981surface}, the boundary term was written in terms of the curvature in the full $D$ dimensions. In \ref{app:Bunch}, we have shown that this expression, when written in terms of the $(D-1)$-dimensional boundary curvature tensor, reduces to our expression above. The conjugate momentum, the coefficient of $\df h^{dq}$ above, was previously derived by Davis \cite{Davis:2002gn} and also by Gravanis and Willison \cite{Gravanis:2002wy} (also see \cite{Deruelle:2003ck} where it is explicitly written down). In \ref{app:GB:Conj:Mom}, we have shown that our conjugate momentum matches with these expressions. The total derivative term, to our knowledge, has not been explicitly provided in the literature before.
\subsection{Manipulating the Boundary Term for general Lanczos-Lovelock: Part 2}\label{sec:BT_LL-2}

In this section, we shall continue the derivation of the boundary variation for general \LL gravity from where we left off in \ref{sec:BT_LL-1}. We had simplified the boundary variation for general \LL gravity, leaving out the $\sqh$ factor, to the form given in \ref{BT_LL_1}. This form is reproduced below:
\begin{align}
\mathcal{B}[n_{c}]=D_{b}\left( B^{qbd}\delta h_{dq}-\epsilon A^{b}_{q}\delta u^{q} \right)-2\epsilon A^{d}_{m}\delta K^{m}_{d}+\left(
D^{b} B_{qbd}+\epsilon A^{b}_{d}K _{bq}\right)\delta h^{dq} \label{BT_LL_2}~.
\end{align}
The term to be beat into shape in the above expression is $-2\epsilon A^{d}_{m}\delta K^{m}_{d}$, which shall now be ruthlessly decomposed. We use \ref{def_A} and \ref{P_def} to write
\begin{align}
-2\epsilon A^{d}_{m}\delta K^{m}_{d}&=-4\epsilon P_{uv}^{wx}n_{w}n^{v}h^{u}_{m}h^{d}_{x}\delta K^{m}_{d}
\nonumber
\\
&=-4\epsilon n_{w}n^{v}h^{u}_{e}h^{f}_{x}\delta K^{e}_{f}\sum _{m}c_{m}\frac{m}{2^{m}}\delta ^{wxc_{1}d_{1}\cdots c_{m-1}d_{m-1}}_{uva_{1}b_{1}\cdots a_{m-1}b_{m-1}}R^{a_{1}b_{1}}_{c_{1}d_{1}}\cdots R^{a_{m-1}b_{m-1}}_{c_{m-1}d_{m-1}}~.
\end{align}
Here, the determinant tensor has indices $w$ and $v$ contracted with the normal. This means that only the surface components of the Riemann tensor factors $R^{a_{i}b_{i}}_{c_{i}d_{i}}$ do not contribute. For example, consider decomposing the index $a_1$ in $R^{a_{1}b_{1}}_{c_{1}d_{1}}$ into its surface and normal components. We have
\begin{equation}
R^{a_{1}b_{1}}_{c_{1}d_{1}}=\delta^{a_{1}}_{e} R^{e b_{1}}_{c_{1}d_{1}}=\left(h^{a_{1}}_{e}+\epsilon n^{a_{1}}n_{e}\right) R^{e b_{1}}_{c_{1}d_{1}}~.
\end{equation}
The factor $n^{a_{1}}$ together with $n^{v}$ will get killed because of the antisymmetry of the determinant tensor. So only the surface component, $h^{a_{1}}_{e} R^{e b_{1}}_{c_{1}d_{1}}$ survives. Since this is true for all indices of all the Riemann factors, we can write
\begin{align}
-2 \epsilon A^{d}_{m}\delta K^{m}_{d}=&-4\epsilon n_{w}n^{v}h^{u}_{e}h^{f}_{x}\delta K^{e}_{f} \nn \\
&\times\left[\sum _{m}c_{m}\frac{m}{2^{m}}\delta ^{wxc_{1}d_{1}\cdots c_{m-1}d_{m-1}}_{uva_{1}b_{1}\cdots a_{m-1}b_{m-1}}\left(h^{a_{1}}_{p}h^{b_{1}}_{q}h^{r}_{c_{1}}h^{s}_{d_{1}}R^{pq}_{rs}\right)
\cdots \left(h^{a_{m-1}}_{i}h^{b_{m-1}}_{j}h^{k}_{c_{m-1}}h^{l}_{d_{m-1}}R^{ij}_{kl}\right)\right]~.
\end{align}
Looking at the definition of the determinant tensor from \ref{determinant_tensor}, we can see that the the above term will have various contractions of the form $n_a h^a_b$ which will all reduce to zero. In fact, the only terms which will survive are the ones which have contraction of $n_{w}$ with $n^{v}$. Writing $\delta ^{wxc_{1}d_{1}\cdots c_{m-1}d_{m-1}}_{uva_{1}b_{1}\cdots a_{m-1}b_{m-1}}$ as $-\delta ^{wxc_{1}d_{1}\cdots c_{m-1}d_{m-1}}_{vua_{1}b_{1}\cdots a_{m-1}b_{m-1}}$ and using \ref{determinant_tensor}, we have the surviving terms as
\begin{align}
-2 \epsilon A^{d}_{m}\delta K^{m}_{d}=& ~4 h^{u}_{e}h^{f}_{x}\delta K^{e}_{f}\nn\\&\times\left[\sum _{m}c_{m}\frac{m}{2^{m}}\delta ^{xc_{1}d_{1}\cdots c_{m-1}d_{m-1}}_{ua_{1}b_{1}\cdots a_{m-1}b_{m-1}}\left(h^{a_{1}}_{p}h^{b_{1}}_{q}h^{r}_{c_{1}}h^{s}_{d_{1}}R^{pq}_{rs}\right)
\cdots \left(h^{a_{m-1}}_{i}h^{b_{m-1}}_{j}h^{k}_{c_{m-1}}h^{l}_{d_{m-1}}R^{ij}_{kl}\right)\right]~.\label{AK-1}
\end{align}
We can decompose $h^{u}_{e}h^{f}_{x}\delta K^{e}_{f}$ as
\begin{align}
h^{u}_{e}h^{f}_{x}\delta K^{e}_{f}=\delta K^{u}_{x}-K^{e}_{x}\delta h^{u}_{e}-K^{u}_{f}\delta h^{f}_{x}
=\delta K^{u}_{x}+\epsilon K^{u}_{f}n_{x}\delta n^{f}~.
\end{align}
When this expression is substituted in \ref{AK-1}, the structure of the determinant tensor would mean that the $n_x$ in the second term would either get contracted with one of the factors of $h$ or with $K^{u}_{f}$. Hence, this term will get killed and one obtains
\begin{align}
-2\epsilon A^{d}_{m}\delta K^{m}_{d}&=4 \sum _{m}c_{m}\frac{m}{2^{m}}\delta ^{vc_{1}d_{1}\cdots c_{m-1}d_{m-1}}_{ua_{1}b_{1}\cdots a_{m-1}b_{m-1}}\left(h^{a_{1}}_{p}h^{b_{1}}_{q}h^{r}_{c_{1}}h^{s}_{d_{1}}R^{pq}_{rs}\right)
\cdots \left(h^{a_{m-1}}_{i}h^{b_{m-1}}_{j}h^{k}_{c_{m-1}}h^{l}_{d_{m-1}}R^{ij}_{kl}\right)\delta K^{u}_{v}~.
\end{align}
Using the Gauss-Codazzi relation, the above expression becomes
\begin{align}
-2 \epsilon A^{d}_{m}\delta K^{m}_{d}&= \sum _{m} c_{m}\frac{m}{2^{m-2}}\delta ^{vc_{1}d_{1}\cdots c_{m-1}d_{m-1}}_{ua_{1}b_{1}\cdots a_{m-1}b_{m-1}}\left[\mathcal{R}^{a_{1}b_{1}}_{c_{1}d_{1}}-\epsilon 
\left(K^{a_{1}}_{c_{1}}K^{b_{1}}_{d_{1}}- K^{a_{1}}_{d_{1}}K^{b_{1}}_{c_{1}}\right)\right]
\nonumber
\\
&\phantom{=4 \sum _{m}}\cdots \left[\mathcal{R}^{a_{m-1}b_{m-1}}_{c_{m-1}d_{m-1}}-\epsilon 
\left(K^{a_{m-1}}_{c_{m-1}}K^{b_{m-1}}_{d_{m-1}}- K^{a_{m-1}}_{d_{m-1}}K^{b_{m-1}}_{c_{m-1}}\right)\right]\delta K^{u}_{v}
\nonumber
\\
&= \sum _{m} c_{m}\frac{m}{2^{m-2}}\delta ^{vc_{1}d_{1}\cdots c_{m-1}d_{m-1}}_{ua_{1}b_{1}\cdots a_{m-1}b_{m-1}}\left[\mathcal{R}^{a_{1}b_{1}}_{c_{1}d_{1}}-2\epsilon K^{a_{1}}_{c_{1}}K^{b_{1}}_{d_{1}}\right]
\nonumber
\\
&\phantom{=4 \sum _{m}}\cdots \left[\mathcal{R}^{a_{m-1}b_{m-1}}_{c_{m-1}d_{m-1}}-2\epsilon K^{a_{m-1}}_{c_{m-1}}K^{b_{m-1}}_{d_{m-1}}\right]\delta K^{u}_{v}
\nonumber
\\
&= \sum _{m}\left[c_{m}\frac{m}{2^{m-2}}\delta ^{vc_{1}d_{1}\cdots c_{m-1}d_{m-1}}_{ua_{1}b_{1}\cdots a_{m-1}b_{m-1}} \phantom{R^{a_{s}b_{s}}_{c_{s}d_{s}}K^{a_{s+1}}_{c_{s+1}}}\phantom{\sum _{s=0}^{m-1}\Mycomb[m-1]{s}~\epsilon ^{m-1-s}\mathcal{R}^{a_{1}b_{1}}_{c_{1}d_{1}}}\right. \nn\\
&\phantom{=4 \sum _{m}\left[\right]}\left.\sum _{s=0}^{m-1}\Mycomb[m-1]{s}~(-2\epsilon)^{m-1-s}\mathcal{R}^{a_{1}b_{1}}_{c_{1}d_{1}}\cdots \mathcal{R}^{a_{s}b_{s}}_{c_{s}d_{s}}K^{a_{s+1}}_{c_{s+1}}K^{b_{s+1}}_{d_{s+1}}\cdots K^{a_{m-1}}_{c_{m-1}}K^{b_{m-1}}_{d_{m-1}}\right]\delta K^{u}_{v} \nn\\
&= \sum _{m}\left[c_{m}\frac{m}{2^{m-2}}\delta ^{vc_{1}d_{1}\cdots c_{m-1}d_{m-1}}_{ua_{1}b_{1}\cdots a_{m-1}b_{m-1}} \phantom{R^{a_{s}b_{s}}_{c_{s}d_{s}}K^{a_{s+1}}_{c_{s+1}}}\phantom{\sum _{s=0}^{m-1}\Mycomb[m-1]{s}~\epsilon ^{m-1-s}\mathcal{R}^{a_{1}b_{1}}_{c_{1}d_{1}}}\right. \nn\\
&\phantom{=4 \sum _{m}\left[\right]}\left.\sum _{s=0}^{m-1}\Mycomb[m-1]{s}~(-2\epsilon)^{m-1-s}\left(\prod_{i=1}^{s}\mathcal{R}^{a_{i}b_{i}}_{c_{i}d_{i}}\right)\left(\prod_{j=s+1}^{m-1}K^{a_{j}}_{c_{j}}K^{b_{j}}_{d_{j}}\right)\right]\delta K^{u}_{v}~.
\end{align}
In the second-last line above, it is to be understood that the indices are to be taken only from the pool $(a_1,b_1,c_1,d_1)$ to $(a_{m-1},b_{m-1},c_{m-1},d_{m-1})$. For example, $s=0$ would give $\mathcal{R}^{a_{1}b_{1}}_{c_{1}d_{1}}\cdots \mathcal{R}^{a_{0}b_{0}}_{c_{0}d_{0}}$ which is to be taken to mean that there are no $\mathcal{R}$ factors in this term. This is clear from the way the expression is written in the last line above. To see how the final expression is arrived at, first note that on multiplying the factors of the form $\left[\mathcal{R}^{a_{i}b_{i}}_{c_{i}d_{i}}-2\epsilon K^{a_{i}}_{c_i}K^{b_i}_{d_{i}}\right]$, we will get a sum of terms each having $(m-1)$ factors, with each set of indices in $\left(a_{i},b_{i},c_{i},d_{i}\right)$, $i=1,...,m-1$, represented by either $R^{a_{i}b_{i}}_{c_{i}d_{i}}$ or $-2\epsilon K^{a_{i}}_{c_i}K^{b_i}_{d_{i}}$. All the terms with $s$ factors of $R^{a_{i}b_{i}}_{c_{i}d_{i}}$ are equal due to the properties of the determinant tensor. The number of terms 
with $s$ 
factors of $R^{a_{i}b_{i}}_{c_{i}d_{i}}$ is $\Mycomb[m-1]{s}$, the number of ways of choosing $s$ factors out of the $(m-1)$ factors of $\left[\mathcal{R}^{a_{i}b_{i}}_{c_{i}d_{i}}-2\epsilon K^{a_{i}}_{c_i}K^{b_i}_{d_{i}}\right]$ to provide $R^{a_{i}b_{i}}_{c_{i}d_{i}}$. The above expression can be written by taking all the $K's$ inside the $\delta$ as 
\begin{align}
-2 \epsilon A^{d}_{m}\delta K^{m}_{d}
&= \sum _{m}\left\{\frac{c_{m}m}{2^{m-2}}\delta ^{vc_{1}d_{1}\cdots c_{m-1}d_{m-1}}_{ua_{1}b_{1}\cdots a_{m-1}b_{m-1}}
\left.\sum _{s=0}^{m-1}\left[\Mycomb[m-1]{s}~(-2\epsilon)^{m-1-s}\mathcal{R}^{a_{1}b_{1}}_{c_{1}d_{1}}\cdots \mathcal{R}^{a_{s}b_{s}}_{c_{s}d_{s}}\phantom{\frac{\delta \left(K^{a_{s+1}}_{c_{s+1}}K^{b_{s+1}}_{d_{s+1}}\right)}{2(m-s)-1}} \right.\right.\right.\nn\\ &\phantom{= \sum _{m}\left\{c_{m}\frac{m}{2^{m-2}}\delta ^{vc_{1}d_{1}\cdots c_{m-1}d_{m-1}}_{ua_{1}b_{1}\cdots a_{m-1}b_{m-1}}\sum _{s=0}^{m-1}\left[\right]\right.}\left.\left.\frac{\delta \left(K^{a_{s+1}}_{c_{s+1}}K^{b_{s+1}}_{d_{s+1}}\cdots K^{a_{m-1}}_{c_{m-1}}K^{b_{m-1}}_{d_{m-1}}K^{u}_{v}\right)}{2(m-s)-1}\right]\right\}~,
\end{align}
since the indices $(u,v)$ can be exchanged with the indices on any of the $K^{a_{i}}_{c_{i}}$ factors with the generation of two minus signs which cancel to give no net sign change. Taking $\delta$ commonly outside, we get the total variation term and term with variation of the boundary Ricci tensor:
\begin{align}
-2 \epsilon A^{d}_{m}\delta K^{m}_{d}
&= \sum _{m}\delta\left\{\frac{c_{m}m}{2^{m-2}}\delta ^{vc_{1}d_{1}\cdots c_{m-1}d_{m-1}}_{ua_{1}b_{1}\cdots a_{m-1}b_{m-1}}
\left.\sum _{s=0}^{m-1}\left[\Mycomb[m-1]{s}~(-2\epsilon)^{m-1-s}\mathcal{R}^{a_{1}b_{1}}_{c_{1}d_{1}}\cdots \mathcal{R}^{a_{s}b_{s}}_{c_{s}d_{s}}\phantom{\frac{\delta \left(K^{a_{s+1}}_{c_{s+1}}K^{b_{s+1}}_{d_{s+1}}\right)}{2(m-s)-1}} \right.\right.\right.\nn\\ &\phantom{= \sum _{m}\left\{c_{m}\frac{m}{2^{m-2}}\delta ^{vc_{1}d_{1}\cdots c_{m-1}d_{m-1}}_{ua_{1}b_{1}\cdots a_{m-1}b_{m-1}}\sum _{s=0}^{m-1}\left[\right]\right.}\left.\left.\frac{ \left(K^{a_{s+1}}_{c_{s+1}}K^{b_{s+1}}_{d_{s+1}}\cdots K^{a_{m-1}}_{c_{m-1}}K^{b_{m-1}}_{d_{m-1}}K^{u}_{v}\right)}{2(m-s)-1}\right]\right\} \nn\\
&\phantom{=}- \sum _{m}\left\{\frac{c_{m}m}{2^{m-2}}\delta ^{vc_{1}d_{1}\cdots c_{m-1}d_{m-1}}_{ua_{1}b_{1}\cdots a_{m-1}b_{m-1}}
\left.\sum _{s=1}^{m-1}\left[s\Mycomb[m-1]{s}~(-2\epsilon)^{m-1-s}\mathcal{R}^{a_{1}b_{1}}_{c_{1}d_{1}}\cdots \mathcal{R}^{a_{s-1}b_{s-1}}_{c_{s-1}d_{s-1}} \phantom{\frac{\delta \left(K^{a_{s+1}}_{c_{s+1}}K^{b_{s+1}}_{d_{s+1}}\right)}{2(m-s)-1}} \right.\right.\right.\nn\\ &\phantom{= \sum _{m}\left\{c_{m}\frac{m}{2^{m-2}}\delta ^{vc_{1}d_{1}\cdots c_{m-1}d_{m-1}}_{ua_{1}b_{1}\cdots a_{m-1}b_{m-1}}\sum _{s=0}^{m-1}\right.}\left.\left.\frac{ \left(K^{a_{s+1}}_{c_{s+1}}K^{b_{s+1}}_{d_{s+1}}\cdots K^{a_{m-1}}_{c_{m-1}}K^{b_{m-1}}_{d_{m-1}}K^{u}_{v}\right)}{2(m-s)-1}\delta\left(\mathcal{R}^{a_{s}b_{s}}_{c_{s}d_{s}}\right)\right]\right\}~.\label{hold}
\end{align}
Note that the second sum in the above expression only runs from $s=1$ since the $s=0$ term does not have any $\mathcal{R}^{a_{s}b_{s}}_{c_{s}d_{s}}$ factor. In the last step above, we have used the fact that terms generated by $\delta$ acting on any of the $\mathcal{R}^{a_{i}b_{i}}_{c_{i}d_{i}}$ are all equivalent due to the properties of the determinant tensor. The $\delta\left(\mathcal{R}^{a_{s}b_{s}}_{c_{s}d_{s}}\right)$ term does not contain any normal derivatives of the variations of the metric (as will be clear when we expand and simplify this expression shortly). Hence, the total variation in the expression above, once the $\sqh$ factor is taken inside, is the term to be canceled by adding an additional boundary term.

Before manipulating the $\delta\left(\mathcal{R}^{a_{s}b_{s}}_{c_{s}d_{s}}\right)$ term in \ref{hold}, we write the above expression in condensed form by introducing the notations
\begin{align}
BT_m\equiv&-\left\{\frac{m}{2^{m-2}}\delta ^{vc_{1}d_{1}\cdots c_{m-1}d_{m-1}}_{ua_{1}b_{1}\cdots a_{m-1}b_{m-1}}
\sum _{s=0}^{m-1}\left[\Mycomb[m-1]{s}~(-2\epsilon)^{m-1-s}\left(\prod_{i=1}^{s}\mathcal{R}^{a_{i}b_{i}}_{c_{i}d_{i}}\right) \frac{ K^{u}_{v} \left(\prod_{j=s+1}^{m-1}K^{a_{j}}_{c_{j}}K^{b_{j}}_{d_{j}}\right)}{2(m-s)-1}\right]\right\}\label{BTm}
\end{align}
and
\begin{align}
S_{a_{s}b_{s}\ (m,s)}^{c_{s}d_{s}}\equiv\frac{m}{2^{m-2}}\delta ^{vc_{1}d_{1}\cdots c_{s}d_{s} \cdots c_{m-1}d_{m-1}}_{ua_{1}b_{1}\cdots a_{s}b_{s} \cdots a_{m-1}b_{m-1}}
s\Mycomb[m-1]{s}~(-2\epsilon)^{m-1-s}\left(\prod_{i=1}^{s-1}\mathcal{R}^{a_{i}b_{i}}_{c_{i}d_{i}}\right) \frac{ K^{u}_{v} \left(\prod_{j=s+1}^{m-1}K^{a_{j}}_{c_{j}}K^{b_{j}}_{d_{j}}\right)}{2(m-s)-1}~,\label{S-def}
\end{align}
where $s$ in the last expression can take values from $1$ to $m-1$.
Then \ref{hold} can be written as
\begin{align}
-2 \epsilon A^{d}_{m}\delta K^{m}_{d}
&=c_{m} \left\{-\sum _{m}\delta\left(BT_m\right)- \sum _{m}\sum _{s=1}^{m-1}\left[S_{a_{s}b_{s}\ (m,s)}^{c_{s}d_{s}}\delta\left(\mathcal{R}^{a_{s}b_{s}}_{c_{s}d_{s}}\right)\right]\right\}~.\label{hold-1}
\end{align}
To separate out variations of $h^{ab}$ and surface derivative terms from $\delta\left(\mathcal{R}^{a_{s}b_{s}}_{c_{s}d_{s}}\right)$, first we write 
\begin{equation}
\delta\left(\mathcal{R}^{a_{s}b_{s}}_{c_{s}d_{s}}\right)=g^{b_{s}e_{s}}\delta\left(\mathcal{R}^{a_{s}}_{\phantom{a}e_{s}c_{s}d_{s}}\right)+\mathcal{R}^{a_{s}}_{\phantom{a}e_{s}c_{s}d_{s}}\delta g^{b_{s}e_{s}}~.
\end{equation}
Writing $\df g^{b_{s}e_{s}}=\df h^{b_{s}e_{s}}+\epsilon n^{b_s} \df n^{e_s}+\epsilon  n^{e_s}\df n^{b_s}$, we can see that the components along the normals will get killed in \ref{hold-1} because the determinant tensor will contract the normals either with $K^a_b$'s or with a factor of $\mathcal{R}^{ab}_{cd}=h^{a}_{p}h^{b}_{q}h^{r}_{c}h^{s}_{d}R^{pq}_{rs}+\epsilon 
\left(K^{a}_{c}K^{b}_{d}- K^{a}_{d}K^{b}_{c}\right)$. Thus, the $\delta\left(\mathcal{R}^{a_{s}b_{s}}_{c_{s}d_{s}}\right)$ term in \ref{hold-1} becomes
\begin{align}
- \sum _{m}\sum _{s=1}^{m-1}\left\{S_{a_{s}b_{s}\ (m,s)}^{c_{s}d_{s}}\delta\left(\mathcal{R}^{a_{s}b_{s}}_{c_{s}d_{s}}\right)\right\}&=- \sum _{m}\sum _{s=1}^{m-1}\left\{S_{a_{s}b_{s}\ (m,s)}^{c_{s}d_{s}}\left[g^{b_{s}e_{s}}\delta\left(\mathcal{R}^{a_{s}}_{\phantom{a}e_{s}c_{s}d_{s}}\right)+\mathcal{R}^{a_{s}}_{\phantom{a}e_{s}c_{s}d_{s}}\delta h^{b_{s}e_{s}}\right]\right\}\nn\\&=- \sum _{m}\sum _{s=1}^{m-1}\left\{S_{a_{s}\ (m,s)}^{\phantom{a_{s}}e_{s}c_{s}d_{s}}\delta\left(\mathcal{R}^{a_{s}}_{\phantom{a}e_{s}c_{s}d_{s}}\right)+S_{a_{s}b_{s}\ (m,s)}^{c_{s}d_{s}}\mathcal{R}^{a_{s}}_{\phantom{a}e_{s}c_{s}d_{s}}\delta h^{b_{s}e_{s}}\right\}~. \label{hold-2}
\end{align}
Next we shall decompose $\delta\left(\mathcal{R}^{a_{s}}_{\phantom{a}e_{s}c_{s}d_{s}}\right)$. In BNC (see \ref{app:coord}), we have the following relation for the boundary Riemann tensor:
\begin{align}
\delta\left(\mathcal{R}^{\alpha_{s}}_{\phantom{a}\epsilon_{s}\gamma_{s}\delta_{s}}\right)=D_{\gamma_{s}}\df\gamma^{\alpha_{s}}_{\epsilon_{s}\delta_{s}}-D_{\delta_{s}}\df\gamma^{\alpha_{s}}_{\epsilon_{s}\gamma_{s}}
\end{align}
Using the prescription in \ref{app:bulk-boundary_correspondence}, this implies the following tensorial relation between the bulk quantities:
\begin{align}
\delta\left(\mathcal{R}^{a_{s}}_{\phantom{a}e_{s}c_{s}d_{s}}\right)=D_{c_{s}}\df\overline{\gamma}^{a_{s}}_{e_{s}d_{s}}-D_{d_{s}}\df\overline{\gamma}^{a_{s}}_{e_{s}c_{s}},
\end{align}
where the object $\df\overline{\gamma}^{a}_{bc}$ was defined in \ref{3gamma_1_u}. Thus, 
\begin{align}
S_{a_{s}\ (m,s)}^{\phantom{a_{s}}e_{s}c_{s}d_{s}}\delta\left(\mathcal{R}^{a_{s}}_{\phantom{a}e_{s}c_{s}d_{s}}\right)=S_{a_{s}\ (m,s)}^{\phantom{a_{s}}e_{s}c_{s}d_{s}} \left(D_{c_{s}}\df\overline{\gamma}^{a_{s}}_{e_{s}d_{s}}-D_{d_{s}}\df\overline{\gamma}^{a_{s}}_{e_{s}c_{s}}\right)~.
\end{align}  
The indices $c_s$ and $d_s$ in $S_{a_{s}\ (m,s)}^{\phantom{a_{s}}e_{s}c_{s}d_{s}}$ are antisymmetric due to properties of the determinant tensor. So we can write
\begin{align}
S_{a_{s}\ (m,s)}^{\phantom{a_{s}}e_{s}c_{s}d_{s}}\delta\left(\mathcal{R}^{a_{s}}_{\phantom{a}e_{s}c_{s}d_{s}}\right)=2S_{a_{s}\ (m,s)}^{\phantom{a_{s}}e_{s}c_{s}d_{s}} \left(D_{c_{s}}\df\overline{\gamma}^{a_{s}}_{e_{s}d_{s}}\right)~.\label{SR-0}
\end{align}
We would like to take $D_{c_{s}}$ commonly outside to get a total derivative. But, $S_{a_{s}\ (m,s)}^{\phantom{a_{s}}e_{s}c_{s}d_{s}}$ is not a purely boundary tensor since contraction of any index with the normal vector will not give zero (see the Gauss-Bonnet case evaluated in \ref{S_Gauss_Bonnet_pre}). But only the purely boundary part of this tensor will contribute since 
\begin{align}
S_{a_{s}\ (m,s)}^{\phantom{a_{s}}e_{s}c_{s}d_{s}}\delta\left(\mathcal{R}^{a_{s}}_{\phantom{a}e_{s}c_{s}d_{s}}\right)=2S_{a_{s}\ (m,s)}^{\phantom{a_{s}}e_{s}c_{s}d_{s}} \left(D_{c_{s}}\df\overline{\gamma}^{a_{s}}_{e_{s}d_{s}}\right)=2S_{a_{s}\ (m,s)}^{\phantom{a_{s}}e_{s}c_{s}d_{s}} h^{f_{s}}_{c_{s}} h^{a_{s}}_{g_{s}}h^{i_{s}}_{e_{s}}h^{j_{s}}_{d_{s}}\left(D_{f_{s}}\df\overline{\gamma}^{g_{s}}_{i_{s}j_{s}}\right)~.\label{SR-01}
\end{align}
Thus, defining 
\begin{align} \label{S_tilde}
\widetilde{S}_{g_{s}\ (m,s)}^{\phantom{a_{s}}i_{s}f_{s}j_{s}}=S_{a_{s}\ (m,s)}^{\phantom{a_{s}}e_{s}c_{s}d_{s}} h^{f_{s}}_{c_{s}} h^{a_{s}}_{g_{s}}h^{i_{s}}_{e_{s}}h^{j_{s}}_{d_{s}},
\end{align}
we can write
\begin{align}
S_{a_{s}\ (m,s)}^{\phantom{a_{s}}e_{s}c_{s}d_{s}}\delta\left(\mathcal{R}^{a_{s}}_{\phantom{a}e_{s}c_{s}d_{s}}\right)=2\widetilde{S}_{a_{s}\ (m,s)}^{\phantom{a_{s}}e_{s}c_{s}d_{s}} \left(D_{c_{s}}\df\overline{\gamma}^{a_{s}}_{e_{s}d_{s}}\right)=2D_{c_{s}}\left(\widetilde{S}_{a_{s}\ (m,s)}^{\phantom{a_{s}}e_{s}c_{s}d_{s}} \df\overline{\gamma}^{a_{s}}_{e_{s}d_{s}}\right)-2D_{c_{s}}\left(\widetilde{S}_{a_{s}\ (m,s)}^{\phantom{a_{s}}e_{s}c_{s}d_{s}}\right) \df\overline{\gamma}^{a_{s}}_{e_{s}d_{s}}~.\label{SR-1}
\end{align}
Using \ref{3gamma_1_u}, the second term in \ref{SR-1} can be written as
\begin{align}
\nl-2D_{c_{s}}\left(\widetilde{S}_{a_{s}\ (m,s)}^{\phantom{a_{s}}e_{s}c_{s}d_{s}}\right) \df\overline{\gamma}^{a_{s}}_{e_{s}d_{s}}\nn\\&=-D_{c_{s}}\left(\widetilde{S}_{a_{s}\ (m,s)}^{\phantom{a_{s}}e_{s}c_{s}d_{s}}\right) \left\{h^{a_s f_s}\left[-D_{f_s} \left(h^{i_s}_{e_s}h^{j_s}_{d_s} \df h_{i_s j_s}\right)+D_{e_s} \left(h^{i_s}_{d_s}h^{j_s}_{f_s} \df h_{i_s j_s}\right)+D_{d_s} \left(h^{i_s}_{e_s}h^{j_s}_{f_s} \df h_{i_s j_s}\right)\right]\right\} \nn\\
&=-D_{c_{s}}\left(\widetilde{S}^{f_{s}e_{s}c_{s}d_{s}}_{(m,s)}\right) \left[-D_{f_{s}} \left(h^{i_s}_{e_s}h^{j_s}_{d_s} \df h_{i_s j_s}\right)+D_{e_s} \left(h^{i_s}_{d_s}h^{j_s}_{f_s} \df h_{i_s j_s}\right)+D_{d_s} \left(h^{i_s}_{e_s}h^{j_s}_{f_s} \df h_{i_s j_s}\right)\right] \nn \\
&=2 D_{c_{s}}\left(\widetilde{S}^{f_{s}e_{s}c_{s}d_{s}}_{(m,s)}\right) D_{f_{s}} \left(h^{i_s}_{e_s}h^{j_s}_{d_s} \df h_{i_s j_s}\right) \nn\\
&=2 D_{f_{s}} \left[D_{c_{s}}\left(\widetilde{S}^{f_{s}e_{s}c_{s}d_{s}}_{(m,s)}\right) \left(h^{i_s}_{e_s}h^{j_s}_{d_s} \df h_{i_s j_s}\right)\right] -2 \left(D_{f_{s}} D_{c_{s}}\widetilde{S}^{f_{s}e_{s}c_{s}d_{s}}_{(m,s)}\right)\left(h^{i_s}_{e_s}h^{j_s}_{d_s} \df h_{i_s j_s}\right)\nn\\
&=2 D_{f_{s}} \left[D_{c_{s}}\left(\widetilde{S}^{f_{s}e_{s}c_{s}d_{s}}_{(m,s)}\right)\df h_{e_s d_s}\right] -2 \left(D_{f_{s}} D_{c_{s}}\widetilde{S}^{f_{s}e_{s}c_{s}d_{s}}_{(m,s)}\right)\df h_{e_s d_s}~,
\end{align}
where we have made use of the antisymmetry of indices $f_s$ and $e_s$ in $\widetilde{S}^{f_{s}e_{s}c_{s}d_{s}}_{(m,s)}$. Substituting back in \ref{SR-1}, we obtain
\begin{align}
&S_{a_{s}\ (m,s)}^{\phantom{a_{s}}e_{s}c_{s}d_{s}}\delta\left(\mathcal{R}^{a_{s}}_{\phantom{a}e_{s}c_{s}d_{s}}\right)=2D_{c_{s}}\left[\widetilde{S}_{a_{s}\ (m,s)}^{\phantom{a_{s}}e_{s}c_{s}d_{s}} \df\overline{\gamma}^{a_{s}}_{e_{s}d_{s}}+D_{f_{s}}\left(\widetilde{S}^{c_{s}e_{s}f_{s}d_{s}}_{(m,s)}\right) \delta h_{e_s d_s}\right] -2 \left(D_{f_{s}} D_{c_{s}}\widetilde{S}^{f_{s}e_{s}c_{s}d_{s}}_{(m,s)}\right) \delta h_{e_s d_s}\nn\\
&=2D_{c_{s}}\left[\widetilde{S}_{a_{s}\ (m,s)}^{\phantom{a_{s}}e_{s}c_{s}d_{s}} \df\overline{\gamma}^{a_{s}}_{e_{s}d_{s}}-D^{f_{s}}\left(\widetilde{S}^{c_{s}\phantom{e_{s}}\phantom{f_{s}}\phantom{d_{s}}}_{\phantom{c_{s}}e_{s}f_{s}d_{s}\ (m,s)}\right) \delta h^{e_s d_s}\right] +2 \left(D_{f_{s}} D^{c_{s}}\widetilde{S}^{f_{s}}_{\phantom{a_s}e_{s}c_{s}d_{s}\ (m,s)}\right) \delta h^{e_s d_s} \label{dR-final}
~.
\end{align}
Putting \ref{dR-final} in \ref{hold-2}, we obtain
\begin{align}
&- \sum _{m}\sum _{s=1}^{m-1}\left\{S_{a_{s}b_{s}\ (m,s)}^{c_{s}d_{s}}\delta\left(\mathcal{R}^{a_{s}b_{s}}_{c_{s}d_{s}}\right)\right\}=- \sum _{m}\sum _{s=1}^{m-1}\left\{S_{a_{s}b_{s}\ (m,s)}^{c_{s}d_{s}}\left[g^{b_{s}e_{s}}\delta\left(\mathcal{R}^{a_{s}}_{\phantom{a}e_{s}c_{s}d_{s}}\right)+\mathcal{R}^{a_{s}}_{\phantom{a}e_{s}c_{s}d_{s}}\delta h^{b_{s}e_{s}}\right]\right\}\nn\\&=- \sum _{m}\sum _{s=1}^{m-1}\left\{2D_{c_{s}}\left[\widetilde{S}_{a_{s}\ (m,s)}^{\phantom{a_{s}}e_{s}c_{s}d_{s}} \df\overline{\gamma}^{a_{s}}_{e_{s}d_{s}}-D^{f_{s}}\left(\widetilde{S}^{c_{s}\phantom{e_{s}}\phantom{f_{s}}\phantom{d_{s}}}_{\phantom{c_{s}}e_{s}f_{s}d_{s}\ (m,s)}\right) \delta h^{e_s d_s}\right] +2 \left(D_{f_{s}} D^{c_{s}}\widetilde{S}^{f_{s}}_{\phantom{a_s}e_{s}c_{s}d_{s}\ (m,s)}\right) \delta h^{e_s d_s}\right.\nn\\
&\phantom{=-\sum _{m}\sum _{s=0}^{m-1}\left(\right)}\left.+S_{a_{s}b_{s}\ (m,s)}^{c_{s}d_{s}}\mathcal{R}^{a_{s}}_{\phantom{a}e_{s}c_{s}d_{s}}\delta h^{b_{s}e_{s}}\right\}\nn\\&=- \sum _{m}\sum _{s=1}^{m-1}\left\{2D_{c_{s}}\left[\widetilde{S}_{a_{s}\ (m,s)}^{\phantom{a_{s}}e_{s}c_{s}d_{s}} \df\overline{\gamma}^{a_{s}}_{e_{s}d_{s}}-D^{f_{s}}\left(\widetilde{S}^{c_{s}\phantom{e_{s}}\phantom{f_{s}}\phantom{d_{s}}}_{\phantom{c_{s}}e_{s}f_{s}d_{s}\ (m,s)}\right) \delta h^{e_s d_s}\right] +2 \left(D_{f_{s}} D^{c_{s}}\widetilde{S}^{f_{s}}_{\phantom{a_s}e_{s}c_{s}d_{s}\ (m,s)}\right) \delta h^{e_s d_s}\right.\nn\\
&\phantom{=-\sum _{m}\sum _{s=0}^{m-1}\left(\right)}\left.+\widetilde{S}_{a_{s}b_{s}\ (m,s)}^{c_{s}d_{s}}\mathcal{R}^{a_{s}}_{\phantom{a}e_{s}c_{s}d_{s}}\delta h^{b_{s}e_{s}}\right\}~, \label{hold-3}
\end{align}
since $\mathcal{R}^{a_{s}}_{\phantom{a}e_{s}c_{s}d_{s}}$ is orthogonal to the normal in all indices and $n_{b_s}\delta h^{b_{s}e_{s}}=0$ as $\df n_a \propto n_a$.
Substituting this in \ref{hold-1},
\begin{align}
-2 \epsilon A^{d}_{m}\delta K^{m}_{d}
= &- \sum _{m} c_m\left\{\delta\left(BT_m\right)+ \sum _{s=1}^{m-1}\left\{2D_{c_{s}}\left[\widetilde{S}_{a_{s}\ (m,s)}^{\phantom{a_{s}}e_{s}c_{s}d_{s}} \df\overline{\gamma}^{a_{s}}_{e_{s}d_{s}}-D^{f_{s}}\left(\widetilde{S}^{c_{s}\phantom{e_{s}}\phantom{f_{s}}\phantom{d_{s}}}_{\phantom{c_{s}}e_{s}f_{s}d_{s}\ (m,s)}\right) \delta h^{e_s d_s}\right]\right. \right.\nn\\
&\phantom{=- c_m\left\{\right.\delta\left(BT_m\right)-\sum _{m}}\left.\phantom{\sum _{s=1}^{m-1}}\left.+ \left(2D_{f_{s}} D^{c_{s}}\widetilde{S}^{f_{s}}_{\phantom{a_s}e_{s}c_{s}b_{s}\ (m,s)}+\widetilde{S}_{a_{s}b_{s}\ (m,s)}^{c_{s}d_{s}}\mathcal{R}^{a_{s}}_{\phantom{a}e_{s}c_{s}d_{s}}\right) \delta h^{e_s b_s}\right\}\right\}~.\label{hold-11}
\end{align}
and \ref{hold} in \ref{BT_LL_1}, we obtain
\begin{align}
\mathcal{B}[n_{c}]&=D_{b}\left( B^{qbd}\delta h_{dq}-\epsilon A^{b}_{q}\delta u^{q} \right)- \sum _{m}c_m\sum _{s=1}^{m-1}\left\{2D_{c_{s}}\left[\widetilde{S}_{a_{s}\ (m,s)}^{\phantom{a_{s}}e_{s}c_{s}d_{s}} \df\overline{\gamma}^{a_{s}}_{e_{s}d_{s}}-D^{f_{s}}\left(\widetilde{S}^{c_{s}\phantom{e_{s}}\phantom{f_{s}}\phantom{d_{s}}}_{\phantom{c_{s}}e_{s}f_{s}d_{s}\ (m,s)}\right) \delta h^{e_s d_s}\right]\right\} \nn\\
&\phantom{=}-\sum_m c_m \df\left(BT_m\right)\nn\\&\phantom{=} +\left(
D^{b} B_{qbd}+\epsilon A^{b}_{d}K _{bq}\right)\delta h^{dq}  - \sum _{m}c_m\sum _{s=1}^{m-1}\left\{\left(2D_{f_{s}} D^{c_{s}}\widetilde{S}^{f_{s}}_{\phantom{a_s}e_{s}c_{s}b_{s}}+\widetilde{S}_{a_{s}b_{s}\ (m,s)}^{c_{s}d_{s}}\mathcal{R}^{a_{s}}_{\phantom{a}e_{s}c_{s}d_{s}}\right) \delta h^{e_s b_s}\right\} \label{BT_LL_prefinal}~.
\end{align}
The full boundary variation term can be obtained by adding the $\sqh$ factor and writing
\begin{align}
&\sqh\mathcal{B}[n_{c}] \nn\\
&=\sqh \left\{D_{b}\left( B^{qbd}\delta h_{dq}-\epsilon A^{b}_{q}\delta u^{q} \right)- \sum _{m}c_m \sum _{s=1}^{m-1}\left\{2D_{c_{s}}\left[ \widetilde{S}_{a_{s}\ (m,s)}^{\phantom{a_{s}}e_{s}c_{s}d_{s}} \df\overline{\gamma}^{a_{s}}_{e_{s}d_{s}}- D^{f_{s}}\left(\widetilde{S}^{c_{s}\phantom{e_{s}}\phantom{f_{s}}\phantom{d_{s}}}_{\phantom{c_{s}}e_{s}f_{s}d_{s}\ (m,s)}\right) \delta h^{e_s d_s}\right]\right\}\right\} \nn\\
&\phantom{=}-\sum_m c_m  \df\left(\sqh BT_m\right)\nn\\
&\phantom{=} +\sqh\left(
D^{b} B_{qbd}+\epsilon A^{b}_{d}K _{bq}-\frac{1}{2}\sum_m  BT_m h_{dq}\right)\delta h^{dq}\nn\\&\phantom{=} - \sqh\sum _{m}c_m\sum _{s=1}^{m-1}\left\{\left(2D_{f_{s}} D^{c_{s}}\widetilde{S}^{f_{s}}_{\phantom{a_s}e_{s}c_{s}b_{s}}+\widetilde{S}_{a_{s}b_{s}\ (m,s)}^{c_{s}d_{s}}\mathcal{R}^{a_{s}}_{\phantom{a}e_{s}c_{s}d_{s}}\right) \delta h^{e_s b_s}\right\} \label{BT_LL_final}~,
\end{align}
with $BT_m$ defined in \ref{BTm}, $\df \overline{\gamma}^a_{bc}$ defined in \ref{3gamma_1_u} and $\widetilde{S}_{a_{s}\ (m,s)}^{\phantom{a_{s}}e_{s}c_{s}d_{s}}$ defined in \ref{S_tilde} ($S_{a_{s}\ (m,s)}^{\phantom{a_{s}}e_{s}c_{s}d_{s}}$ is defined in \ref{S-def}). The first line above has the total derivative on the surface, the second line is the total variation term to be canceled by adding a boundary term and the last two lines give the contribution to be killed by fixing the boundary metric. 

Having derived the full structure of the boundary term for Lanczos-Lovelock gravity, we will briefly comment on the Brown-York formulation of quasi-local energy in this context. The equivalent of $-2K\sqrt{h}$ for Lanczos-Lovelock gravity is given in the second line of \ref{BT_LL_final} and allows us to define the corresponding quasi-local energy for Lanczos-Lovelock gravity, in terms of an integral of this expression over a co-dimension two surface. This would match with the result obtained in \cite{Chakraborty:2015kva} since they have used the expressions for boundary terms existing in the literature \cite{Miskovic:2007mg,Padmanabhan:2013xyr}, and we will show below that the boundary term that we have derived matches with these expressions. The coefficient of $\delta h_{cd}$ (the \textit{lower} components) will represent the corresponding Brown-York stress tensor for Lanczos-Lovelock gravity, as already mentioned in \cite{Miskovic:2007mg}. These definitions are worthy of further exploration from different perspectives, e.g., as regards the conservation of the Brown-York stress-energy tensor on-shell, physical significance of associated conserved quantities, etc. This will require a detailed study of the $1+(D-1)$ decomposition of Lanczos-Lovelock gravity, which we hope to report on in a future work.

\subsection{Consistency Checks on Our Expressions}\label{sec:consistency}

\subsubsection{Boundary Term}
The boundary term to be added to the action for general \LL theory can be read off from \ref{BT_LL_final} as
\begin{align}
\sqh\mathcal{B}[n_{c}]=\sum_m c_m \sqh BT_m,
\end{align}
with $BT_m$ defined in \ref{BTm} and reproduced below:
\begin{align}
BT_m=&-\left\{\frac{m}{2^{m-2}}\delta ^{vc_{1}d_{1}\cdots c_{m-1}d_{m-1}}_{ua_{1}b_{1}\cdots a_{m-1}b_{m-1}}
\sum _{s=0}^{m-1}\left[\Mycomb[m-1]{s}~(-2\epsilon)^{m-1-s}\left(\prod_{i=1}^{s}\mathcal{R}^{a_{i}b_{i}}_{c_{i}d_{i}}\right) \frac{ K^{u}_{v} \left(\prod_{j=s+1}^{m-1}K^{a_{j}}_{c_{j}}K^{b_{j}}_{d_{j}}\right)}{2(m-s)-1}\right]\right\}.\label{BTm-1}
\end{align}
As the first consistency check, we have verified that $BT_m$ reduces to the correct expressions for Einstein-Hilbert and Gauss-Bonnet for $m=1$ and $m=2$ respectively (see \ref{BT_EH_final} and \ref{GB_BT_final}).

Another consistency check would be to compare with previous expressions in literature. The expression for the boundary term for general \LL has been previously given in the literature \cite{Miskovic:2007mg,Padmanabhan:2013xyr} as
\begin{align}
A_{\rm sur (m)} &= \int \limits_{\partial \mathcal{V}}  d^{(D-1)} x  \; \sqrt{-h} \; C_m
\end{align}
with
\begin{align}
C_m &= 2 m \int \limits_0^1 \DM s \; \delta^{i_1 i_2 i_3... i_{2m-1}}_{j_1 j_2 j_3...
	j_{2m-1}} \; 
K^{j_1}_{i_1} \left( \frac{\mathcal{R}^{j_2 j_3}_{i_2 i_3}}{2}-s^2 K^{j_2}_{i_2}
K^{j_3}_{i_3}\right)
\ldots
\left(\frac{\mathcal{R}^{j_{2m-2} j_{2m-1}}_{i_{2m-2} i_{2m-1}}}{2}-s^2
K^{j_{2m-2}}_{i_{2m-2}} K^{j_{2m-1}}_{i_{2m-1}}\right)~.
\end{align}
In \ref{app:Olea_Misk}, we have shown that the expression for $C_m$ reduces to the expression for $BT_m$ given in \ref{BTm-1}. Thus, our results agree with previous results in literature derived in a different manner, affirming the consistency of our methods.
\subsubsection{Surface Derivative Term and Term with Variation of Boundary Metric}

We are not aware of any previous occurrences of these terms in the literature for general Lovelock gravity. So, as consistency checks, we shall evaluate these terms for $m=1$ and $m=2$ and compare them with the Einstein-Hilbert and \GB expressions derived earlier. (We would like to note that all the Einstein-Hilbert and \GB expressions except for the total derivative term in the \GB case have been provided in the literature and match with the expressions derived here, see \ref{app:GB:Conj:Mom}.)  We have performed the relevant calculations in \ref{app:GB_EH_from_LL} and confirmed that the general expressions do reduce to the corresponding Einstein-Hilbert and Gauss-Bonnet expressions. 
\section{Results}

In this paper, we have separated out the boundary variation for a general \LL theory into a surface total derivative term, the term to be canceled by the addition of a boundary term to the action and the Dirichlet variation term which is to be put to zero by fixing the intrinsic metric at the boundary. \textit{As far as we know, the surface total derivative term and the Dirichlet variation term  have not been explicitly presented in any of the previous literature, for the general Lanczos-Lovelock model.} The boundary variation for general \LL is the integral over the boundary of
\begin{align}
&\sqh\mathcal{B}[n_{c}] \nn\\
&=\sqh \left\{D_{b}\left( B^{qbd}\delta h_{dq}-\epsilon A^{b}_{q}\delta u^{q} \right)- \sum _{m}c_m \sum _{s=1}^{m-1}\left\{2D_{c_{s}}\left[ \widetilde{S}_{a_{s}\ (m,s)}^{\phantom{a_{s}}e_{s}c_{s}d_{s}} \df\overline{\gamma}^{a_{s}}_{e_{s}d_{s}}- D^{f_{s}}\left(\widetilde{S}^{c_{s}\phantom{e_{s}}\phantom{f_{s}}\phantom{d_{s}}}_{\phantom{c_{s}}e_{s}f_{s}d_{s}\ (m,s)}\right) \delta h^{e_s d_s}\right]\right\}\right\} \nn\\
&\phantom{=}-\sum_m c_m  \df\left(\sqh BT_m\right)\nn\\
&\phantom{=} +\sqh\left(
D^{b} B_{qbd}+\epsilon A^{b}_{d}K _{bq}-\frac{1}{2}\sum_m  BT_m h_{dq}\right)\delta h^{dq}\nn\\&\phantom{=} - \sqh\sum _{m}c_m\sum _{s=1}^{m-1}\left\{\left(2D_{f_{s}} D^{c_{s}}\widetilde{S}^{f_{s}}_{\phantom{a_s}e_{s}c_{s}b_{s}}+\widetilde{S}_{a_{s}b_{s}\ (m,s)}^{c_{s}d_{s}}\mathcal{R}^{a_{s}}_{\phantom{a}e_{s}c_{s}d_{s}}\right) \delta h^{e_s b_s}\right\} \label{BT_LL_final_results}~,
\end{align} 
where $B^{qbd}$ is defined in \ref{def_B}, $A^{b}_{q}$ is defined in \ref{def_A}, $\df u^{q}=\df n^{q}+g^{qd}\df n_d$, $\widetilde{S}_{a_{s}\ (m,s)}^{\phantom{a_{s}}e_{s}c_{s}d_{s}}$ is defined in \ref{S_tilde}, $\df \overline{\gamma}^a_{bc}$ is defined in \ref{3gamma_1_u}, $\mathcal{R}^{a_{s}}_{\phantom{a}e_{s}c_{s}d_{s}}$ is the boundary Riemann tensor and the boundary term to be added to the action is the integral over the boundary of $\sqh\sum_m c_m BT_m$ with
\begin{align}
BT_m\equiv&-\left\{\frac{m}{2^{m-2}}\delta ^{vc_{1}d_{1}\cdots c_{m-1}d_{m-1}}_{ua_{1}b_{1}\cdots a_{m-1}b_{m-1}}
\sum _{s=0}^{m-1}\left[\Mycomb[m-1]{s}~(-2\epsilon)^{m-1-s}\left(\prod_{i=1}^{s}\mathcal{R}^{a_{i}b_{i}}_{c_{i}d_{i}}\right) \frac{ K^{u}_{v} \left(\prod_{j=s+1}^{m-1}K^{a_{j}}_{c_{j}}K^{b_{j}}_{d_{j}}\right)}{2(m-s)-1}\right]\right\}\label{BTm-result}~.
\end{align}
We have verified that this result matches with previous results in the literature \cite{Miskovic:2007mg,Yale:2011dq}.
We have also separately derived the Gauss-Bonnet case to obtain the boundary variation as integral over the boundary of
\begin{align}
\nl\sqh\mathcal{B}[n_{c}]\nn\\=& \sqh D_{b}\left\lbrace B^{qbd}_{\ (2)}\delta h_{dq}-\epsilon A^{b}_{q\ (2)}\delta u^{q}+\left(S^{pq}\delta \overline{\gamma} ^{b}_{pq}-S^{ab}\delta \overline{\gamma} ^{c}_{ac}\right)-\left[D^{d}S^{bc}-\frac{1}{2}\left(D^{b}S^{cd}+h^{cd}D_{a}S^{ab}\right) \right]\delta h_{cd}\right\rbrace
\nonumber
\\
&-\delta\left\{\sqh BT_2\right\}
\nonumber
\\
&+\sqh\Big[D^{b} B_{qbd\ (2)}+\epsilon A^{b}_{d\ (2)}K _{bq}+8 \left(K^{m}_{d}\mathcal{R}_{mq}-\frac{K}{2}\mathcal{R}_{dq}\right)\nn\\
&\phantom{+\sqh\Big[]}+ \frac{1}{2} \left(h_{dq}D_{a} D_{b}S^{ab}+D_{c}D^{c}S_{dq}\right) - D_{c}D_{d}S^{c}_{q}-\frac{1}{2}BT_2 h_{dq}
\Big]\delta h^{dq}~,\label{GB_BT_final_sqh_result}
\end{align} 
where $B_{qbd\ (2)}$ is defined in \ref{def_B_m}, $A^{b}_{d\ (2)}$ is defined in \ref{def_A_m}, $S^{ab}=8(K^{ab}-(1/2)Kh^{ab})$ and the boundary term to be added to the action is the integral over the boundary of
\begin{equation}\label{GB_boundary_term_result}
\sqh BT_2 = \sqh \left[8 \mathcal{R}^{a}_{b}K^{b}_{a}-4 \mathcal{R}K+\epsilon\left(\frac{4}{3}K^{3}+\frac{8}{3}K^{a}_{b}K^{b}_{c}K^{c}_{a}-4KK^{a}_{b}K^{b}_{a}\right)\right]~.
\end{equation}
This result matches previous results given in \cite{bunch1981surface,Yale:2011dq}, while the conjugate momentum, the coefficient of $\df h^{dq}$ above, matches the result obtained in \cite{Davis:2002gn,Gravanis:2002wy} (explicit expression provided in \cite{Deruelle:2003ck}). \textit{As far as we know, the total derivative term has not been explicitly presented in the previous literature.}

Finally, we have also verified that our general \LL expressions reduce to the corresponding \EH and \GB expressions for $m=1$ and $m=2$.
\section*{Acknowledgements}

Research of TP is partially supported by  J. C. Bose Research Grant, DST, Government of India. KP would like to thank Perimeter Institute for Theoretical Physics, Waterloo, Canada for kind hospitality during the early part of this work and Dean Carmi for discussions that led to the initiation of this work. Research of S.C. is supported by the SERB-NPDF grant (PDF/2016/001589) from DST, Government of India.
\appendix
\labelformat{section}{Appendix #1}
\labelformat{subsection}{Appendix #1}
\labelformat{subsubsection}{Appendix #1}
\section*{Appendices}

\section{Adapted Coordinate Systems} \label{app:coord}

While dealing with the boundary and quantities defined on the boundary, it is useful to work in a coordinate system where the boundary is given by the constant value of one of the coordinates. In $D$ dimensions, let us define the coordinates $\left(\phi,x^1,\cdots,x^{D-1} \right)$ such that the boundary is given by $\phi=$constant. We shall use Greek letters for indices running over the boundary coordinates. Let us call these boundary adapted coordinates (BAC), which will turn out to be good for our calculations. This coordinate system is good because we have the normal one-form as
\begin{align}
n_a = A \nabla_a \phi =(A,0,0,0),
\end{align}  
where $A$ is the normalization factor. Hence, any tensor $T^{a\cdots b}_{c \cdots d}$ with an upper index $a$ orthogonal to $n_a$, $T^{a\cdots b}_{c \cdots d}n_a=0$, will have $T^{\phi\cdots b}_{c \cdots d}=0$. In other words, that index will only run over the boundary coordinates. For example, a vector $V^a$ that lies on the boundary surface, i.e. satisfies $V^a n_a=0$, will be represented by $V^{a}=\left(0,V^{\alpha}\right)$.

We can go one-up on this useful coordinate system by defining a coordinate system which we shall call boundary adapted normal coordinates (BNC), which are Gaussian normal coordinates \cite{gravitation,MTW} erected near the boundary surface so that the metric appears as 
\begin{equation}
ds^2 = \epsilon d\phi^2 + h_{\alpha \beta} dx^{\alpha}dx^{\beta},
\end{equation}
where $\epsilon$ is $-1$ or $+1$ depending on whether the boundary is spacelike or timelike, respectively. In the case of these coordinates, we have the further property that the normal vector
\begin{align}
n^a = \nabla^a \phi =(1,0,0,0),
\end{align} 
where the normalization factor $A$ is now set to one since $g_{\phi \phi}=\epsilon$. (One may also set it to $\epsilon$ or $-\epsilon$ depending on which way one wants the normal to point.) In this coordinate system, any index of a tensor $T^{a\cdots b}_{c \cdots d}$ orthogonal to the normal will run only over the boundary coordinates and hence may be replaced by Greek indices in any summation. For example, consider the object $h_{ab}dx^a dx^b$ with the induced metric $h_{ab}=g_{ab}-\epsilon n_a n_b$. This is a scalar and hence is the same in any coordinate system. In BNC, we have $h_{ab}dx^a dx^b=h_{\alpha \beta}dx^{\alpha} dx^{\beta}$, from which it is easy to see that this object is the line element on the boundary surface. Thus, $h_{ab}dx^a dx^b$ is the line element on the boundary surface (in any coordinate system).  
\section{Upgrading Boundary Relations to Bulk Relations}\label{app:bulk-boundary_correspondence}

One can use the boundary adapted normal coordinates (BNC) defined above to convert relations of Riemannian geometry valid on the boundary to relations on the bulk. In BNC, consider some tensorial relation,
\begin{align} \label{TS_bound}
{}_{(D-1)}T^{\alpha \beta \cdots}_{\gamma \delta \cdots}={}_{(D-1)}S^{\alpha \beta \cdots}_{\gamma \delta \cdots}~,
\end{align}
between two tensors which are intrinsic to the boundary and functions of the boundary metric $h_{\alpha \beta}$, its variations $\df h_{\alpha \beta}$ and their boundary derivatives $\partial_\alpha$ or boundary covariant derivatives $D_\alpha$. 
One may now introduce two objects in the bulk by functional continuation of the tensors ${}_{(D-1)}T^{\alpha \beta \cdots}_{\gamma \delta \cdots}$ and ${}_{(D-1)}S^{\alpha \beta \cdots}_{\gamma \delta \cdots}$ into the bulk, i.e. we shall construct them to be of the same functional form but with $h_{\alpha \beta}$ replaced by $h_{ab}=g_{ab}-\epsilon n_a n_b$ (likewise for $h^{\alpha \beta}$), $\df h^{\alpha \beta}$ replaced by $\df h^{ab}$, all the derivatives $\partial_\alpha$ upgraded to projected bulk derivatives $h^{b}_a \partial_b$ and covariant derivatives $D_{\alpha}$ upgraded to $D_{a}$ (for example, $D_\alpha V^\beta$ is to be upgraded to $D_a V^b=h^c_a h^b_d \nabla_c V^d$). One has to be careful with $\df h_{\alpha \beta}$ since $\df h_{ab}$ is not orthogonal to the normal even though $\df h^{ab}$ is. This is because $n_a\df h^{ab}=-h^{ab}\df n_a=0$ as $\df n_a \propto n_a$, while $\df n^a$ is not in general proportional to $n^a$. Thus, the right quantity to replace $\df h_{ab}$ is not $\df h_{ab}$ 
but $h^c_a h^d_b\df h_{cd}$, which is seen to be orthogonal to the normal in both indices 
and reduces to $\df h_{\alpha \beta}$ in BNC. While upgrading $D_{\alpha}$ to $D_{a}$, one should make sure that the quantity it acts upon is orthogonal to the boundary in all its indices since $D_{a}$ is only defined as a surface covariant derivative on such objects \cite{gravitation}. But this will happen automatically for the type of tensors we have considered if the procedure outlined above is followed. For example, $D_{c} \df h_{ab}$ is not the right bulk quantity corresponding to $D_{\gamma} \df h_{\alpha \beta}$, but $D_{c}\left(h^d_a h^e_b\df h_{de}\right)$, which is what we obtain by our procedure, would work.    

It is not right now clear to us that the quantities so obtained will be tensors if they contain partial derivatives instead of covariant derivatives. Let us assume that we can look at the resulting expressions and convince ourselves that they are tensors. Then, we shall denote these tensors by the symbols ${}_{(D)}T^{a b \cdots}_{c d \cdots}$ and ${}_{(D)}S^{a b \cdots}_{c d \cdots}$. The indices that appear outside in the tensors ${}_{(D-1)}T^{\alpha \beta \cdots}_{\gamma \delta \cdots}$ and ${}_{(D-1)}S^{\alpha \beta \cdots}_{\gamma \delta \cdots}$ will be orthogonal to the normal since they will come from some form of the induced metric $h_{ab}$ (even $D_a$ is defined by projecting ordinary covariant derivative with the induced metric). The question in front of us now is whether
\begin{align}\label{TS_bulk}
{}_{(D)}T^{a b \cdots}_{c d \cdots}\overset{?}{=}{}_{(D)}S^{a b \cdots}_{c d \cdots}~
\end{align}
is a valid bulk relation. In fact, it is. Since this is a tensorial relation, its validity in any one frame ensures its validity in general. In BNC, one can see that $h_{ab}$ has only the components $h_{\alpha \beta}$ (similarly for $h^{ab}$), $h^{\alpha}_{\beta}$ is $\df^{\alpha}_{\beta}$, $\df h^{ab}$ is $\df h^{\alpha \beta}$, $h^{b}_a \partial_b$ will only have the components $\partial_\alpha$ and any covariant derivative $D_{a}\left(B^{bc\cdots}_{de\cdots}\right)=D_{\alpha}\left(B^{\beta \gamma\cdots}_{\delta \epsilon\cdots}\right)$. Finally, $h^c_a h^d_b\df h_{cd}=\df h_{\alpha \beta}$. Thus, \ref{TS_bulk} reduces to \ref{TS_bound} in BNC since the above reductions are just the opposite of the functional continuation we did to obtain the bulk quantities from the boundary quantities. Since we have assumed \ref{TS_bound} to be valid, \ref{TS_bulk} is valid in BNC and, being a tensorial relation, is also valid in general. 

As an illustration, we have the following relation valid from the view of a bug living on the boundary:  
\begin{equation}
\delta\left[\mathcal{R}_{\alpha \beta}\right]=\left(D_{\mu}\delta \gamma ^{\mu}_{\alpha \beta}-D_{\beta}\delta \gamma ^{\mu}_{\alpha \mu}\right)~.
\end{equation}
By using our procedure, the above equation can be written in terms of bulk quantities as
\begin{align}
\delta\left[\mathcal{R}_{a b}\right]=\left(D_{m}\delta \overline{\gamma} ^{m}_{a b}-D_{b}\delta \overline{\gamma} ^{m}_{a m}\right)~,
\end{align}
where $\mathcal{R}_{a b}$ and $\delta \overline{\gamma}^{m}_{a b}$ (see \ref{3gamma_1_u}) are obtained by taking their formulas in terms of $h_{\alpha \beta}$ and its derivatives and then upgrading using the prescription described above.

We can also go one step further and consider cases where the tensors ${}_{(D)}T^{a b \cdots}_{c d \cdots}$ and ${}_{(D)}S^{a b \cdots}_{c d \cdots}$ contain some other quantities, extrinsic curvature $K^{ab}$ for example, that are orthogonal to the normal in all its indices. Even in that case, we can use the above route if we can prove that the relations with the boundary components of these other quantities ($K^{\alpha \beta}$ for extrinsic curvature) are valid in BNC. 
\section{$\delta \gamma^{a}_{bc}$ Under Coordinate Transformations} \label{app:gamma_gamma_bar}
Consider the quantity:
\begin{equation}
\gamma^a_{bc}=\frac{h^{ad}}{2}\left(-\partial_d h_{bc}+\partial_b h_{cd}+\partial_{c}h_{bd}\right)~.
\end{equation}
This may be taken as the extension of the boundary connection to the bulk. In any boundary adapted coordinate system (BAC) (see \ref{app:coord}), the boundary components are seen to be
\begin{align}
\gamma^\alpha_{\beta \gamma}&=\frac{h^{\alpha d}}{2}\left(-\partial_d h_{\beta \gamma}+\partial_\beta h_{\gamma d}+\partial_{\gamma}h_{\beta d}\right) \nn\\
&=\frac{h^{\alpha \delta}}{2}\left(-\partial_\delta h_{\beta \gamma}+\partial_\beta h_{\gamma\delta}+\partial_{\gamma}h_{\beta \delta}\right)~,
\end{align}  
which are in fact the components of the boundary connection. 

But the variations of $\gamma^a_{bc}$ under variations of the metric can be seen to be not a tensor, unlike the case of $\delta \Gamma^{a}_{bc}$. The variation of the bulk connection $\Gamma^{a}_{bc}$ is a tensor. This is because $\Gamma^{a}_{bc}$ transforms as
\begin{equation}
\Gamma'^{a}_{bc} = \Lambda^{a'}_{d}\Lambda^{e}_{b'}\Lambda^{f}_{c'}\Gamma^{d}_{ef} - \Lambda^{d}_{b'}\Lambda^{e}_{c'}\partial_d\Lambda^{a'}_{e}~, 
\end{equation}
under transformation of coordinates from $x^a$ to $x'^a$ with $\Lambda^{a'}_{b}=\partial x'^a/\partial x^b$ and $\Lambda^{a}_{b'}=\partial x^a/\partial x'^b$. The second term is independent of the metric. Hence, the variation of $\Gamma^{a}_{bc}$ transforms as a tensor under coordinate transformations:
\begin{equation}
\df \Gamma'^{a}_{bc} = \Lambda^{a'}_{d}\Lambda^{e}_{b'}\Lambda^{f}_{c'}\df \Gamma^{d}_{ef}~.
\end{equation}
But the transformation for $\gamma^{a}_{bc}$ is as follows:
\begin{equation}
\gamma'^{a}_{bc} = \Lambda^{a'}_{d}\Lambda^{e}_{b'}\Lambda^{f}_{c'}\gamma^{d}_{ef} + h^e_i \Lambda^{d}_{b'}\Lambda^{a'}_{e}\partial_d\Lambda^{i}_{c'}~.
\end{equation}
The basic difference is that the $h^e_i$ in the second term was a $\df^e_i$ for the case of $\Gamma^{a}_{bc}$ (the minus sign comes about due to the relation $\Lambda^{e}_{c'}\partial_d\Lambda^{a'}_{e}=-\Lambda^{a'}_{e}\partial_d\Lambda^{e}_{c'}$). The second term does not drop off under variations and hence $\df \gamma^{a}_{bc}$ is not a tensor. But note that $\df h^{a}_{b}=-\epsilon \df\left(n^a n_b\right)$ and hence contains only variations of the normal. This means that this term does not contribute if we only vary the boundary metric. Hence, $\df \gamma^{a}_{bc}$ is indeed a tensor if we restrict to variations that vary only the boundary metric. 
\section{Reducing Bunch's Expression for the Gauss-Bonnet Boundary Term} \label{app:Bunch} 

In this appendix, we shall show that the Gauss-Bonnet boundary term provided in \ref{GB_boundary_term} matches with the result previously derived by Bunch \cite{bunch1981surface}.
Bunch provides the boundary term to be added to the Gauss-Bonnet Lagrangian $L_2$ (with the factors as given in \ref{L_Gauss_Bonnet}) as
\begin{equation}\label{Bunch_BT}
\sqh BT_2= \sqh \left(-4 KR + 8 K h^{ab}R_{ab}- 8 K^{ac}h^{bd}R_{abcd} + \frac{8}{3}\epsilon K^3 - 8 \epsilon K K^{ab}K_{ab}+\frac{16}{3}\epsilon K^{a}_b K^b_c K^c_a\right)~.
\end{equation}
Here, we have replaced the Greek indices used by Bunch by Latin indices as per our convention. First, note that the definition of extrinsic curvature used by Bunch differs by a minus sign from the convention that we have used. We have used
\begin{equation} \label{K_ours}
K_{ab}=-h_a^c \nabla_c n_b,
\end{equation}
following the convention used in \cite{gravitation,MTW}. On the other hand, Bunch uses the convention
\begin{equation} \label{K_bunch}
K_{ab}=h_a^c \nabla_c n_b~.
\end{equation}
While this is not explicitly stated in Bunch's paper, it can be easily inferred. For example, the last equation on page L140 in \cite{bunch1981surface} reads
\begin{equation}
K^{a c} K_{a b;c} n^{b}=K^{a c} n^{b} \nabla_c K_{a b} =-K^{a c} K_{a b}K^{b}_{c}~.
\end{equation}
The LHS can be manipulated as
\begin{equation}
K^{a c} n^{b} \nabla_c K_{a b} = \nabla_c\left(K^{a c} K_{a b} n^{b}\right)-K^{a c} K_{a b}\nabla_c n^{b}=-K^{a c} K_{a b}\nabla_c n^{b}=-K^{a c} K_{a b}\left(h^d_c \nabla_d n^{b}\right)~,
\end{equation}
which makes it clear that Bunch has used the definition \ref{K_bunch}. Making the flip in sign in all the extrinsic curvature factors in \ref{Bunch_BT}, we obtain Bunch's result in our convention:
\begin{equation}\label{Bunch_BT_flipped}
BT_2=  4 KR - 8 K h^{ab}R_{ab}+ 8 K^{ac}h^{bd}R_{abcd} - \frac{8}{3}\epsilon K^3 + 8 \epsilon K K^{ab}K_{ab}-\frac{16}{3}\epsilon K^{a}_b K^b_c K^c_a~.
\end{equation}
We shall now work with the curvature terms to write them in terms of the boundary intrinsic curvatures. The first two terms in the bracket above are
\begin{align}
4 KR - 8 K h^{ab}R_{ab}&=8K\left(\frac{R}{2}- h^{ab}R_{ab}\right)=8K\left(\frac{R}{2}-R+\epsilon n^a n^b R_{ab}\right)=8K\left(\epsilon n^a n^b R_{ab}-\frac{R}{2}\right) \nn\\
&=8\epsilon K\left(R_{ab}-g_{ab}\frac{R}{2}\right)n^a n^b=8\epsilon K G_{ab}n^a n^b~.
\end{align}
We have the result (see Chapter 12 in \cite{gravitation})
\begin{equation}
G_{ab}n^a n^b = \frac{1}{2}\left(K^2-K^{mn}K_{mn}-\epsilon\mathcal{R}\right)~.
\end{equation}
Substituting,
\begin{align}\label{first_two}
4 KR - 8 K h^{ab}R_{ab}=4\epsilon K^3-4\epsilon KK^{mn}K_{mn}-4 K\mathcal{R}~.
\end{align}
Next we shall expand out the third term in \ref{Bunch_BT_flipped}.
\begin{align}
8 K^{ac}h^{bd}R_{abcd}=8 K^{ac}h^{bd} h^e_a h^f_c R_{ebfd}~.
\end{align}
From Chapter 12 in \cite{gravitation}, we have the following formula (after adding the correct $\epsilon$ factors):
\begin{align}
 h^e_a h^f_c h^{bd} R_{ebfd}=\mathcal{R}_{ac}-\epsilon \left(K_{ac}K-K_a^b K_{bc}\right)~.
\end{align}
Thus,
\begin{align} \label{third}
8 K^{ac}h^{bd}R_{abcd}= 8 K^{ac}\mathcal{R}_{ac}-8 \epsilon K K^{ac} K_{ac}+8 \epsilon K_a^b K_{b}^{c}K^{a}_{c}~.  
\end{align}
Substituting \ref{first_two} and \ref{third} in \ref{Bunch_BT_flipped}, we obtain
\begin{align}
BT_2=&  4\epsilon K^3-4\epsilon KK^{mn}K_{mn}-4 K\mathcal{R}+ 8 K^{ac}\mathcal{R}_{ac}-8 \epsilon K K^{ac} K_{ac}+8 \epsilon K_a^b K_{b}^{c}K^{a}_{c} \nn \\
 &\phantom{\sqh \left(\right)}- \frac{8}{3}\epsilon K^3 + 8 \epsilon K K^{ab}K_{ab}-\frac{16}{3}\epsilon K^{a}_b K^b_c K^c_a\nn\\
 =& -4 K\mathcal{R}+ 8 K^{ac}\mathcal{R}_{ac}+\epsilon\left(\frac{4}{3} K^3-4 KK^{mn}K_{mn}+\frac{8}{3}  K_a^b K_{b}^{c}K^{a}_{c}\right) ~.
\end{align}
This matches with our result in \ref{GB_boundary_term}. 
\section{Comparing \LL Boundary Term with Previous Literature}\label{app:Olea_Misk}

In this appendix, we shall show that the boundary term we have derived in \ref{BT_LL_2} is consistent with the boundary term previously provided in the literature in \cite{Miskovic:2007mg,Yale:2011dq}. While the expressions in both these references are essentially the same, we shall find it easier to compare with \cite{Yale:2011dq} since his expressions for the action matches ours with the correct factors and all (see our Einstein-Hilbert and Gauss-Bonnet terms in \ref{L_EH} and \ref{L_Gauss_Bonnet}), while the reference \cite{Miskovic:2007mg} has an extra $-(D-2p)!$ sticking around in the $p-$th order Lovelock term. The boundary term for general \LL is given in \cite{Yale:2011dq} as the integral over the boundary of $\sqh$ times
\begin{equation}
 Q_m =  2m  \int_0^1 dt \delta^{a_1 a_2 \cdots a_{2m-1}}_{b_1 b_2 \cdots b_{2m-1}} K^{b_1}_{a_1} \left( \frac{1}{2} R^{b_2 b_3}_{a_2 a_3} - t^2 K^{b_2}_{a_2} K^{b_3}_{a_3} \right) \cdots \left( \frac{1}{2} R^{b_{2m-2} b_{2m-1}}_{a_{2m-2} a_{2m-1}} - t^2 K^{b_{2m-2}}_{a_{2m-2}} K^{b_{2m-1}}_{a_{2m-1}} \right)~.
\end{equation} 
This $Q_m$ is what we have called $BT_m$, with the corresponding expression provided in \ref{BTm-1} as
\begin{align}
BT_m=&-\left\{\frac{m}{2^{m-2}}\delta ^{vc_{1}d_{1}\cdots c_{m-1}d_{m-1}}_{ua_{1}b_{1}\cdots a_{m-1}b_{m-1}}
\sum _{s=0}^{m-1}\left[\Mycomb[m-1]{s}~(-2\epsilon)^{m-1-s}\left(\prod_{i=1}^{s}\mathcal{R}^{a_{i}b_{i}}_{c_{i}d_{i}}\right) \frac{ K^{u}_{v} \left(\prod_{j=s+1}^{m-1}K^{a_{j}}_{c_{j}}K^{b_{j}}_{d_{j}}\right)}{2(m-s)-1}\right]\right\}. \nn\\
=&-\left\{m\delta ^{vc_{1}d_{1}\cdots c_{m-1}d_{m-1}}_{ua_{1}b_{1}\cdots a_{m-1}b_{m-1}}K^{u}_{v}
\sum _{s=0}^{m-1}\left[\frac{\Mycomb[m-1]{s}~(-\epsilon)^{m-1-s}}{2^{s-1}}\left(\prod_{i=1}^{s}\mathcal{R}^{a_{i}b_{i}}_{c_{i}d_{i}}\right) \frac{ \left(\prod_{j=s+1}^{m-1}K^{a_{j}}_{c_{j}}K^{b_{j}}_{d_{j}}\right)}{2(m-s)-1}\right]\right\}.
\label{BTm-2}
\end{align}
It is up to us now to show that these expressions are equal. Before we start off on this momentous journey, we first note that, just as the case in our comparison with Bunch's Gauss-Bonnet expression in \ref{app:Bunch}, our definition for the extrinsic curvature \cite{gravitation} differs from that used in \cite{Yale:2011dq,Miskovic:2007mg} by a minus sign. This is immediately clear if we look at the case of Einstein-Hilbert. In our convention, a positive $\sqg R$ requires a $-2 \sqh K$ integrated at the boundary to keep it well-defined; while in the case of the other pretenders, a positive $\sqg R$ has as consort a Gibbons-Hawking-York term defined as boundary integral of $2 \sqh K$. We shall see that once this relative minus sign is taken into consideration our results are in perfect agreement with \cite{Yale:2011dq,Miskovic:2007mg}.

So, we take care of this relative minus sign business by flipping the overall sign in $Q_m$ (note that the term is odd in $K_{ab}$). 
\begin{equation} \label{Qm_fixed}
Q_m =  -2m  \int_0^1 dt \delta^{a_1 a_2 \cdots a_{2m-1}}_{b_1 b_2 \cdots b_{2m-1}} K^{b_1}_{a_1} \left( \frac{1}{2}\mathcal{R}^{b_2 b_3}_{a_2 a_3} - t^2 K^{b_2}_{a_2} K^{b_3}_{a_3} \right) \cdots \left( \frac{1}{2}\mathcal{R}^{b_{2m-2} b_{2m-1}}_{a_{2m-2} a_{2m-1}} - t^2 K^{b_{2m-2}}_{a_{2m-2}} K^{b_{2m-1}}_{a_{2m-1}} \right)~.
\end{equation}   
We will now manipulate this expression to bring it to the form of \ref{BTm-2}. First, we forcefully evict all those terms which have no business of staying within the integral sign:
\begin{equation} \label{Qm_fixed-1}
Q_m =  -2m \delta^{a_1 a_2 \cdots a_{2m-1}}_{b_1 b_2 \cdots b_{2m-1}} K^{b_1}_{a_1} \int_0^1 dt  \left( \frac{1}{2}\mathcal{R}^{b_2 b_3}_{a_2 a_3} - t^2 K^{b_2}_{a_2} K^{b_3}_{a_3} \right) \cdots \left( \frac{1}{2}\mathcal{R}^{b_{2m-2} b_{2m-1}}_{a_{2m-2} a_{2m-1}} - t^2 K^{b_{2m-2}}_{a_{2m-2}} K^{b_{2m-1}}_{a_{2m-1}} \right)~.
\end{equation}
Thus, having put these terms in their place,  we next expand out the product of the $(m-1)$ factors inside the integral sign. 
\begin{align} 
Q_m &=  -2m \delta^{a_1 a_2 \cdots a_{2m-1}}_{b_1 b_2 \cdots b_{2m-1}} K^{b_1}_{a_1} \int_0^1 dt \sum_{s=0}^{m-1} \left[\Mycomb[m-1]{s}  \prod_{i=1}^{s}\left(\frac{1}{2}\mathcal{R}^{b_{2i}b_{2i+1}}_{a_{2i}a_{2i+1}}\right) \prod_{j=s+1}^{m-1} \left(- t^2 K^{b_{2j}}_{a_{2j}} K^{b_{2j+1}}_{a_{2j+1}}\right)\right] \nn \\
&=  -2m \delta^{a_1 a_2 \cdots a_{2m-1}}_{b_1 b_2 \cdots b_{2m-1}} K^{b_1}_{a_1} \sum_{s=0}^{m-1}\left[\frac{\Mycomb[m-1]{s}}{2^s} \prod_{i=1}^{s}\left(\mathcal{R}^{b_{2i}b_{2i+1}}_{a_{2i}a_{2i+1}}\right) \prod_{j=s+1}^{m-1} \left( K^{b_{2j}}_{a_{2j}} K^{b_{2j+1}}_{a_{2j+1}}\right) \int_0^1 dt (- t^2)^{m-s-1}\right] 
\nn \\
&=  -2m \delta^{a_1 a_2 \cdots a_{2m-1}}_{b_1 b_2 \cdots b_{2m-1}} K^{b_1}_{a_1} \sum_{s=0}^{m-1}\left[\frac{\Mycomb[m-1]{s}\left(-1\right)^{m-s-1}}{2^s\left[2(m-s)-1\right]}  \prod_{i=1}^{s}\left(\mathcal{R}^{b_{2i}b_{2i+1}}_{a_{2i}a_{2i+1}}\right) \prod_{j=s+1}^{m-1} \left( K^{b_{2j}}_{a_{2j}} K^{b_{2j+1}}_{a_{2j+1}}\right)\right]
\nn \\
&=  -m \delta^{a_1 a_2 \cdots a_{2m-1}}_{b_1 b_2 \cdots b_{2m-1}} K^{b_1}_{a_1} \sum_{s=0}^{m-1}\left[\frac{\Mycomb[m-1]{s}\left(-1\right)^{m-s-1}}{2^{s-1}\left[2(m-s)-1\right]} \prod_{i=1}^{s}\left(\mathcal{R}^{b_{2i}b_{2i+1}}_{a_{2i}a_{2i+1}}\right) \prod_{j=s+1}^{m-1} \left( K^{b_{2j}}_{a_{2j}} K^{b_{2j+1}}_{a_{2j+1}}\right)\right]
 ~.
\end{align}
Comparing with \ref{BTm-2}, it is seen that this is exactly what we have called $BT_m$. Thus, we have confirmed that our results match with the results previously given in \cite{Yale:2011dq,Miskovic:2007mg}.
\section{Other Terms: \EH and \GB Cases from \LL Expressions} \label{app:GB_EH_from_LL}

In this appendix, we shall evaluate the terms in \ref{BT_LL_final} for $m=1$ and $m=2$ and verify that they reduce to the corresponding \EH and \GB expressions. Leaving out the common $\sqh$ factor, the terms to be evaluated are the surface derivative term given by
\begin{align}
	ST_{(m)}= D_{b}\left( B^{qbd}_{(m)}\delta h_{dq}-\epsilon A^{b}_{q\ (m)}\delta u^{q} \right)-  \sum _{s=1}^{m-1}\left\{2D_{c_{s}}\left[ \widetilde{S}_{a_{s}\ (m,s)}^{\phantom{a_{s}}e_{s}c_{s}d_{s}} \df\overline{\gamma}^{a_{s}}_{e_{s}d_{s}}- D^{f_{s}}\left(\widetilde{S}^{c_{s}\phantom{e_{s}}\phantom{f_{s}}\phantom{d_{s}}}_{\phantom{c_{s}}e_{s}f_{s}d_{s}\ (m,s)}\right) \delta h^{e_s d_s}\right]\right\}, \label{ST}
\end{align}
 and the Dirichlet variation term
 \begin{align}
 D_{(m)}=&\left(
 D^{b} B_{qbd\ (m)}+\epsilon A^{b}_{d\ (m)}K _{bq}-\frac{1}{2}  BT_m h_{dq}\right)\delta h^{dq}\nn\\&\phantom{=} - \sum _{s=1}^{m-1}\left\{\left(2D_{f_{s}} D^{c_{s}}\widetilde{S}^{f_{s}}_{\phantom{a_s}e_{s}c_{s}b_{s}\ (m,s)}+\widetilde{S}_{a_{s}b_{s}\ (m,s)}^{c_{s}d_{s}}\mathcal{R}^{a_{s}}_{\phantom{a}e_{s}c_{s}d_{s}}\right) \delta h^{e_s b_s}\right\}~. \label{DV}
 \end{align}
\subsection{The $ m=1$ \EH case}
For the \EH case, we have from \ref{sec:EH_BT} that
\begin{align}
B_{(1)}^{qbd}=0;~A^{b}_{d\ (1)}=-\epsilon h^{b}_{d}~;~ BT_{1}=-2K.
\end{align}
From \ref{S-def}, $m=1$ means that $S^{ab}_{cd\ (m,s)}$, and hence $\widetilde{S}^{ab}_{cd\ (m,s)}$, does not exist as $s$ is defined to take values from $1$ to $m-1$. 
Thus,
\begin{align}
ST_{(1)}= D_{b}\left(h^{b}_{q}\delta u^{q} \right)= D_{b}\left(\delta u^{b} \right),  
\end{align}
and
\begin{align}
 D_{(1)}=&-\left(K _{ab}-K h_{ab}\right)\delta h^{ab}~.
\end{align}
These expressions match with the corresponding expressions from \ref{sec:EH_BT}.
\subsection{The $ m=2$ \GB case}

For $m=2$, $S^{c_s d_s}_{a_s b_s\ (m,s)}$ has only the $S^{c_s d_s}_{a_s b_s\ (2,1)}$ term which, from \ref{S-def}, is 
\begin{align}
S^{c_{1}d_{1}}_{a_{1}b_{1}\ (2,1)}&=2\delta ^{vc_{1}d_{1}}_{u a_{1}b_{1}}K^{u}_{v}
\nonumber
\\
&=2\left[\delta ^{v}_{u}\left(\delta ^{c_{1}}_{a_{1}}\delta ^{d_{1}}_{b_{1}}-\delta ^{c_{1}}_{b_{1}}\delta ^{d_{1}}_{a_{1}}\right)
-\delta ^{v}_{a_{1}}\left(\delta ^{c_{1}}_{u}\delta ^{d_{1}}_{b_{1}}-\delta ^{c_{1}}_{b_{1}}\delta ^{d_{1}}_{u} \right)+\delta ^{v}_{b_{1}}\left(\delta ^{c_{1}}_{u}\delta ^{d_{1}}_{a_{1}}-\delta ^{c_{1}}_{a_{1}}\delta ^{d_{1}}_{u} \right)\right]K^{u}_{v}
\nonumber
\\
&=2\left[K\left(\delta ^{c_{1}}_{a_{1}}\delta ^{d_{1}}_{b_{1}}-\delta ^{c_{1}}_{b_{1}}\delta ^{d_{1}}_{a_{1}}\right)
-\left(K^{c_{1}}_{a_{1}}\delta ^{d_{1}}_{b_{1}}-\delta ^{c_{1}}_{b_{1}}K^{d_{1}}_{a_{1}}\right)
+\left(K^{c_{1}}_{b_{1}}\delta ^{d_{1}}_{a_{1}}-\delta ^{c_{1}}_{a_{1}}K^{d_{1}}_{b_{1}} \right)\right]~.\label{S_Gauss_Bonnet_pre}
\end{align}
Therefore, $\widetilde{S}^{c_{1}d_{1}}_{a_{1}b_{1}\ (2,1)}$ is
\begin{align}
\widetilde{S}^{c_{1}d_{1}}_{a_{1}b_{1}\ (2,1)}=2\left[K\left(h ^{c_{1}}_{a_{1}}h ^{d_{1}}_{b_{1}}-h ^{c_{1}}_{b_{1}}h ^{d_{1}}_{a_{1}}\right)
-\left(K^{c_{1}}_{a_{1}}h ^{d_{1}}_{b_{1}}-h ^{c_{1}}_{b_{1}}K^{d_{1}}_{a_{1}}\right)
+\left(K^{c_{1}}_{b_{1}}h ^{d_{1}}_{a_{1}}-h ^{c_{1}}_{a_{1}}K^{d_{1}}_{b_{1}} \right)\right]~.\label{S_Gauss_Bonnet}
\end{align}
\subsubsection{The Total Derivative Term}

First, let us calculate the total derivative term to compare with \ref{GB_BT_final}. The first two terms in \ref{ST} occur in the same form in \ref{GB_BT_final}. Thus, we have to evaluate the last two terms, which are of the following form for Gauss-Bonnet:
\begin{align}
- 2D_{c_{1}}\left[ \widetilde{S}_{a_{1}\ (2,1)}^{\phantom{a_{1}}e_{1}c_{1}d_{1}} \df\overline{\gamma}^{a_{1}}_{e_{1}d_{1}}- D^{f_{1}}\left(\widetilde{S}^{c_{1}\phantom{e_{1}}\phantom{f_{1}}\phantom{d_{1}}}_{\phantom{c_{1}}e_{1}f_{1}d_{1}\ (2,1)}\right) \delta h^{e_1 d_1}\right] \label{GB_div_rem}
\end{align}
Let us first simplify $2D_{c_1}\left[D^{f_{1}}\left(\widetilde{S}^{c_{1}\phantom{e_{1}}\phantom{f_{1}}\phantom{d_{1}}}_{\phantom{c_{1}}e_{1}f_{1}d_{1}\ (2,1)}\delta h^{e_1 d_1}\right)\right]$. Denoting this by Term 2,
\begin{align}
&\textrm{Term}~2\nn\\&=D_{c_1} \left\{2D^{f_{1}}\left(\widetilde{S}^{c_{1}\phantom{e_{1}}\phantom{f_{1}}\phantom{d_{1}}}_{\phantom{c_{1}}e_{1}f_{1}d_{1}\ (2,1)}\right) \delta h^{e_1 d_1}\right\}
\nonumber
\\
&=D_{c_1} \left\{4D^{f_{1}}\left[K\left(h^{c_1}_{f_1}h_{e_1 d_1}-h^{c_1}_{d_1}h_{e_1 f_1}\right)-\left(K^{c_1}_{f_1}h_{d_1 e_1}-K_{e_1 f_1}h^{c_1}_{d_1}\right)
+\left(K^{c_1}_{d_1}h_{e_1 f_1}-K_{e_1 d_1}h^{c_1}_{f_1} \right)\right]\delta h^{e_1 d_1} \right\}
\nonumber
\\
&=D_{c_1} \left\{4\left(D^{c_1}K\right)h_{e_1 d_1}\delta h^{e_1 d_1}-4\left(D_{e_1}K\right)\delta h^{e_1 c_1}-4\left(D_{f_1}K^{f_1 c_1}\right)h_{e_1 d_1}\delta h^{e_1 d_1}
+4\left(D_{f_1}K^{f_1}_{e_1}\right)\delta h^{e_1 c_1} \right.
\nonumber
\\
&\phantom{=D_{c_1} }\left.+4\left(D_{e_1}K^{c_1}_{d_1}\right)\delta h^{e_1 d_1}-4\left(D^{c_1}K_{e_1 d_1}\right)\delta h^{e_1 d_1}\right\},
\end{align}
where we have used since $n_{a}\df h^{ab}=0$ (because we have $\df n_a \propto n_a$).

On the other hand, we have the following term from \ref{GB_BT_final}:
\begin{equation}
Y_1=-D_{b}\left\{\left[D^{d}S^{bc}-\frac{1}{2}\left(D^{b}S^{cd}+h^{cd}D_{a}S^{ab}\right) \right]\delta h_{cd}\right\}~.
\end{equation}
We have $\df h_{cd}=\df g_{cd}-\epsilon n_c\df n_{d}-\epsilon n_d\df n_{c}$, of which the terms with the normal are killed by the cofactor which is orthogonal to the boundary, leaving us with $\df g_{cd}=-g_{ce}g_{df}\df g^{ef}=-g_{ce}g_{df}\left(\df h^{ef }+\epsilon n^e\df n^{f}+\epsilon n^f\df n^{e}\right)$, of which again the normal components are killed. Thus, substituting $S^{ab}=8(K^{ab}-(1/2)Kh^{ab})$, we obtain
\begin{align}
Y_1=&D_b\left\{\Big[D_{e}S^{b}_{f}-\frac{1}{2}\left(D^{b}S_{ef}+h_{ef}D_{a}S^{ab}\right)\Big]\delta h^{ef}\right\} \nn\\
 = &D_b\left\{8\left(D_{e}K^{b}_{f}\right)\delta h^{ef}-4\left(D_{e}K\right)\delta h^{eb}-4\left(D^{b}K_{ef}\right)\delta h^{ef}\right.
\nonumber
\\
&\left.\phantom{D_b~}+4\left(D^{b}K\right)h_{ef}\delta h^{ef}-4\left(D_{a}K^{ab}\right)h_{ef}\delta h^{ef}\right\}~.
\end{align}
Thus, we can write
\begin{align}
\textrm{Term}~2&=Y_1+4D_c \left[\left(D^{q}K_{pq}\right)\delta h^{cp}-\left(D_{d}K^{c}_{b}\right)\delta h^{bd}\right]~. \label{Term_2_final}
\end{align}
Let us now consider the other term in \ref{GB_div_rem} which reads
\begin{align}
\textrm{Term}~1&=- 2D_{c_{1}}\left[ \widetilde{S}_{a_{1}\ (2,1)}^{\phantom{a_{1}}e_{1}c_{1}d_{1}} \df\overline{\gamma}^{a_{1}}_{e_{1}d_{1}}\right]\nn\\
&=-4D_{c} \left\{\left[K\left(h^{c}_{a}h^{bd}-h^{bc}h^{d}_{a}\right)-\left(K^{c}_{a}h^{bd}-K^{d}_{a}h^{bc}\right)
+\left(K^{cb}h^{d}_{a}-K^{db}h^{c}_{a}\right)\right]\delta \overline{\gamma} ^{a}_{bd}\right\}
\nonumber
\\
&=4D_c\left\{-Kh^{pq}\delta \overline{\gamma} ^{c}_{pq}+K h^{bc}\delta \overline{\gamma} ^{a}_{ba}+K^{c}_{a}h^{bd}\delta \overline{\gamma} ^{a}_{bd}
-K^{d}_{a}h^{bc}\delta \overline{\gamma} ^{a}_{bd}-K^{cb}\delta \overline{\gamma} ^{a}_{ba}+K^{bd}\delta \overline{\gamma} ^{c}_{bd}\right\}~.
\end{align}
We have to compare this with the following term from \ref{GB_BT_final}:
\begin{align}
D_c\left(S^{pq}\delta \overline{\gamma} ^{c}_{pq}-S^{ac}\delta \overline{\gamma} ^{b}_{ab}\right)=D_c\left\{8K^{pq}\delta \overline{\gamma} ^{c}_{pq}-4Kh^{pq}\delta \overline{\gamma} ^{c}_{pq}
-8K^{ac}\delta \overline{\gamma} ^{b}_{ab}+4Kh^{ac}\delta \overline{\gamma} ^{b}_{ab}\right\}~.
\end{align}
We can see that
\begin{align}
\textrm{Term}~1&-D_c\left(S^{pq}\delta \overline{\gamma} ^{c}_{pq}-S^{ac}\delta \overline{\gamma} ^{b}_{ab}\right)=D_c\left\{-4K^{pq}\delta \overline{\gamma} ^{c}_{pq}
+4K^{ac}\delta \overline{\gamma} ^{b}_{ab}-4K^{d}_{a}h^{cb}\delta \overline{\gamma} ^{a}_{bd}+4K^{c}_{a}h^{bd}\delta \overline{\gamma} ^{a}_{bd}\right\}~. \label{T3-}
\end{align}
Let us simplify the last two terms using \ref{3gamma_1_u}:
\begin{align}
4K^{c}_{a}h^{bd}\delta \overline{\gamma} ^{a}_{bd}&=4K^{c}_{a}h^{pq}\left\{\frac{1}{2}h^{as}\left[-D_{s}\left(h^c_p h^d_q\delta h_{cd}\right)+D_{p}\left(h^c_s h^d_q\delta h_{cd}\right)+D_{q}\left(h^c_s h^d_p\delta h_{cd}\right)\right]\right\}
\nonumber
\\
&=-2K^{cs}h^{pq}\left\{\left[D_{s}\left(h^c_p h^d_q\delta h_{cd}\right)+D_{p}\left(h^c_s h^d_q\delta h_{cd}\right)-D_{q}\left(h^c_s h^d_p\delta h_{cd}\right)\right]-2D_{p}\left(h^c_s h^d_q\delta h_{cd}\right) \right\}
\nonumber
\\
&=-4K^{cs}\delta \overline{\gamma} ^{p}_{sp}+4K^{cs}D^{q}\left(h^c_s h^d_q\delta h_{cd}\right),
\end{align}
and
\begin{align}
-4K^{d}_{a}h^{cb}\delta \overline{\gamma} ^{a}_{bd}&=-4K^{d}_{a}h^{cb}\left[\frac{1}{2}h^{ap}\left(-D_{p}\left(h^x_b h^y_d\delta h_{xy}\right)+D_{b}\left(h^x_p h^y_d\delta h_{xy}\right)+D_{d}\left(h^x_p h^y_b\delta h_{xy}\right)\right) \right]
\nonumber
\\
&=2K^{dp}h^{cb}\left\{\left[-D_{b}\left(h^x_p h^y_d\delta h_{xy}\right)+D_{p}\left(h^x_b h^y_d\delta h_{xy}\right)+D_{d}\left(h^x_p h^y_b\delta h_{xy}\right)\right]-2D_{d}\left(h^x_p h^y_b\delta h_{xy}\right)\right\}
\nonumber
\\
&=4K^{dp}\delta \overline{\gamma} ^{c}_{dp}-4K^{dp}h^{cb}D_{d}\left(h^x_p h^y_b\delta h_{xy}\right)~.
\end{align}
Substitution in \ref{T3-} leads to
\begin{align}
\textrm{Term}~1-D_c\left(S^{pq}\delta \overline{\gamma} ^{c}_{pq}-S^{ac}\delta \overline{\gamma} ^{b}_{ab}\right)=& D_c\left[-4K^{pq}\delta \overline{\gamma} ^{c}_{pq}
+4K^{ac}\delta \overline{\gamma} ^{b}_{ab}-4K^{cs}\delta \overline{\gamma} ^{p}_{sp} \right.
\nonumber
\\
&\left. \phantom{D_c}+4K^{cs}D^{q}\left(h^x_s h^y_q\delta h_{xy}\right)+4K^{dp}\delta \overline{\gamma} ^{c}_{dp}-4K^{dp}h^{cb}D_{d}\left(h^x_p h^y_b\delta h_{xy}\right)\right]
\nonumber
\\
=& D_c\left[4K^{cs}D^{q}\left(h^x_s h^y_q\delta h_{xy}\right)-4K^{dp}h^{cb}D_{d}\left(h^x_p h^y_b\delta h_{xy}\right)\right]~. \label{Term_3_final}
\end{align}
From \ref{Term_2_final} and \ref{Term_3_final}, we obtain
\begin{align}
\nl\textrm{Term}~2+\textrm{Term}~1\nn\\
=&D_c \left\{\Big[D_{d}S^{c}_{b}-\frac{1}{2}\left(D^{c}S_{bd}+h_{bd}D_{a}S^{ac}\right)\Big]\delta h^{bd}
+\left(S^{pq}\delta \overline{\gamma} ^{c}_{pq}-S^{ac}\delta \overline{\gamma} ^{b}_{ab}\right)\right.
\nonumber
\\
&\left.\phantom{\frac{1}{2}D_c}-4\left(D_{d}K^{c}_{b}\right)\delta h^{bd}+\left(4D^{q}K_{pq}\right)\delta h^{cp}-4K^{dp}h^{cb}D_{d}\left(h^x_p h^y_b\delta h_{xy}\right)+4K^{cs}D^{q}\left(h^x_s h^y_q\delta h_{xy}\right)\right\}~. \label{surf_div_inter}
\end{align}
The LHS is the term we obtained by putting $m=2$ in our \LL expression while the first line in the RHS gives the corresponding terms in the \GB expression previously derived. Thus, the terms present in the last line have to vanish. These terms can be simplified as follows:
\begin{align}
\hspace{2em}&\hspace{-2em}-4\left(D_{d}K^{c}_{b}\right)\delta h^{bd}+4D^{q}K_{pq}\delta h^{cp}-4K^{dp}h^{cb}D_{d}\left(h^x_p h^y_b\delta h_{xy}\right)+4K^{cs}D^{q}\left(h^x_s h^y_q\delta h_{xy}\right)
\nonumber
\\
&=-4\left(D_{d}K^{c}_{b}\right)\delta h^{bd}+4D^{q}K_{pq}\delta h^{cp}+4K^{d}_{p}D_{d}\delta h^{pc}-4K^{c}_{s}D_{q}\delta h^{sq}
\nonumber
\\
&=-4D_{d}\left(K^{c}_{b}\delta h^{bd}\right)+4D_{d}\left(K^{d}_{p}\delta h^{cp}\right)
=4D_{d}\left(K^{d}_{p}\delta h^{cp}-K^{c}_{p}\delta h^{pd}\right)~.
\end{align}
In the second step above, we have used the relation
\begin{align}
h^{ac}h^{bd}\df h_{cd}=-\df h^{ab},
\end{align}
which can be proved by going to BNC as indicated in \ref{app:bulk-boundary_correspondence}.
Adding the $D_c$,
\begin{align}
\hspace{2em}&\hspace{-2em} D_{c}\Big[-4\left(D_{d}K^{c}_{b}\right)\delta h^{bd}+4D^{q}K_{pq}\delta h^{cp}-4K^{dp}h^{cb}D_{d}\left(h^x_p h^y_b\delta h_{xy}\right)+4K^{cs}D^{q}\left(h^x_s h^y_q\delta h_{xy}\right)\Big]
\nonumber
\\
&=4D_{c}D_{d}\left(K^{d}_{p}\delta h^{cp}-K^{c}_{p}\delta h^{pd}\right)=4\left(D_{c}D_{d}-D_{d}D_{c}\right)K^{d}_{p}\delta h^{cp}\nn \\ &=4\left(\mathcal{R}^{d}_{~ecd}K^{e}_{p}\delta h^{cp}+\mathcal{R}^{c}_{~ecd}K^{d}_{p}\delta h^{ep}\right)=4\left(-\mathcal{R}_{ec}K^{e}_{p}\delta h^{cp}+\mathcal{R}_{ed}K^{d}_{p}\delta h^{ep}\right)=0,
\end{align}
where we have used
\begin{align}
\left(D_{a}D_{b}-D_{b}D_{a}\right)A^{cd} =\mathcal{R}^c_{~eab} A^{ed} +\mathcal{R}^d_{~eab} A^{ce},
\end{align}
valid for any tensor $A^{a}_{b}$ that is orthogonal to the normal on both indices. Thus, the last line in the RHS of \ref{surf_div_inter} vanishes and we obtain
\begin{align}
\hspace{2em}&\hspace{-2em}-2D_{c_1}\Big[\widetilde{S}_{a_1\ (2,1)}^{~b_1 c_1 d_1}\delta \overline{\gamma} ^{a_1}_{b_1 d_1}-\left(D^{a_1}\widetilde{S}^{c_1}_{~d_1 a_1 b_1\ (2,1)}\right)\delta h^{b_1 d_1}\Big]\nn\\
&=D_{c}\Big\lbrace\Big[D_{d}S^{c}_{b}-\frac{1}{2}\left(D^{c}S_{bd}+h_{bd}D_{a}S^{ac}\right)\Big]\delta h^{bd}
+\left(S^{pq}\delta \overline{\gamma} ^{c}_{pq}-S^{ac}\delta \overline{\gamma} ^{b}_{ab}\right)\Big\rbrace~.
\end{align}
Hence, the total derivative term for general \LL gravity reduces to the previously derived \GB result for $m=2$.
\subsubsection{The Dirichlet Variation Term}

In this section, we shall compare \ref{DV} evaluated for $m=2$,
\begin{align}
D_{(2)}=&\left(
D^{b} B_{qbd\ (2)}+\epsilon A^{b}_{d\ (2)}K _{bq}-\frac{1}{2}  BT_2 h_{dq}\right)\delta h^{dq}\nn\\&\phantom{=} - \left(2D_{f_{1}} D^{c_{1}}\widetilde{S}^{f_{1}}_{\phantom{a_1}e_{1}c_{1}b_{1}\ (2,1)}+\widetilde{S}_{a_{1}b_{1}\ (2,1)}^{c_{1}d_{1}}\mathcal{R}^{a_{1}}_{\phantom{a}e_{1}c_{1}d_{1}}\right) \delta h^{e_1 b_1}~.  \label{LL_m_2}
\end{align}
 with the corresponding expression in \ref{GB_BT_final_sqh}, which is
\begin{align}
D_{GB}=&\Big[D^{b} B_{qbd\ (2)}+\epsilon A^{b}_{d\ (2)}K _{bq}+8 \left(K^{m}_{d}\mathcal{R}_{mq}-\frac{K}{2}\mathcal{R}_{dq}\right)\nn\\
&\phantom{\Big[}+ \frac{1}{2} \left(h_{dq}D_{a} D_{b}S^{ab}+D_{c}D^{c}S_{dq}\right) - D_{c}D_{d}S^{c}_{q}-\frac{1}{2}BT_2 h_{dq}
\Big]\delta h^{dq}~. \label{DGB}
\end{align}
The first two terms above and the last term are present in the same form in \ref{LL_m_2}. We need to check that
\begin{align}
\nl - \left(2D_{f_{1}} D^{c_{1}}\widetilde{S}^{f_{1}}_{\phantom{a_1}e_{1}c_{1}b_{1}\ (2,1)}+\widetilde{S}_{a_{1}b_{1}\ (2,1)}^{c_{1}d_{1}}\mathcal{R}^{a_{1}}_{\phantom{a}e_{1}c_{1}d_{1}}\right) \delta h^{e_1 b_1} \nn\\ \overset{?}{=}&\left[8 \left(K^{m}_{d}\mathcal{R}_{mq}-\frac{K}{2}\mathcal{R}_{dq}\right)\right.\nn\\
&\phantom{\Big[}+ \frac{1}{2} \left(h_{dq}D_{a} D_{b}S^{ab}+D_{c}D^{c}S_{dq}\right) - D_{c}D_{d}S^{c}_{q}
\Big]\delta h^{dq}~.
\end{align}
Let us start by computing
\begin{align}
\textrm{Expression}~1=&-2D_{f_{1}}D^{c_{1}}\widetilde{S}^{f_{1}}_{~e_{1}c_{1}b_{1}\ (2,1)}\delta h^{e_{1}b_{1}}
=-2D_{c}D^{a}\widetilde{S}^{c}_{~dab\ (2,1)}\delta h^{db}
\nonumber
\\
=&-4D_{c}D^{a}\left[K\left(h ^{c}_{a}h_{bd}-h^{c}_{b}h_{ad}\right)-\left(K^{c}_{a}h_{bd}-K_{ad}h^{c}_{b} \right)+\left(K^{c}_{b}h_{ad}-K_{bd}h^{c}_{a}\right)\right]\delta h^{bd}
\nonumber
\\
=&-4\left(D_{a}D^{a}K\right)h_{bd}\delta h^{bd}+4\left(D_{b}D_{d}K\right)\delta h^{bd}+4\left(D_{c}D_{a}K^{ca}\right)h_{bd}\delta h^{bd}
\nonumber
\\
&-4\left(D_{b}D_{a}K^{a}_{d}\right)\delta h^{bd}-4\left(D_{c}D_{d}K^{c}_{b}\right)\delta h^{bd}+4\left(D_{a}D^{a}K_{bd}\right)\delta h^{bd}~.
\end{align}
On the other hand, we can substitute in \ref{DGB} $S^{ab}=8(K^{ab}-(1/2)Kh^{ab})$ and obtain
\begin{align}
\hspace{2em}&\hspace{-2em}\Big[\frac{1}{2}\left(h_{dq}D_{a}D_{b}S^{ab}+D_{c}D^{c}S_{dq}\right)-D_{c}D_{d}S^{c}_{q}\Big]\delta h^{dq} \nn\\
=&4\left(D_{a}D_{b}K^{ab}\right)h_{dq}\delta h^{dq}-4\left(D^{a}D_{a}K\right)h_{dq}\delta h^{dq}
\nonumber
\\
&+4\left(D_{c}D^{c}K_{dq}\right)\delta h^{dq}-8\left(D_{c}D_{d}K^{c}_{q}\right)\delta h^{dq}
+4\left(D_{q}D_{d}K\right)\delta h^{dq},
\end{align}
so that
\begin{align}
\textrm{Expression}~1=&\Big[\frac{1}{2}\left(h_{dq}D_{a}D_{b}S^{ab}+D_{c}D^{c}S_{dq}\right)-D_{c}D_{d}S^{c}_{q}\Big]\delta h^{dq}
\nonumber
\\
&+4\left(D_{c}D_{d}K^{c}_{b}\right)\delta h^{bd}-4\left(D_{b}D^{a}K_{ad}\right)\delta h^{bd}
\nonumber
\\
=&\Big[\frac{1}{2}\left(h_{dq}D_{a}D_{b}S^{ab}+D_{c}D^{c}S_{dq}\right)-D_{c}D_{d}S^{c}_{q}\Big]\delta h^{dq}
\nonumber
\\
&+4\left(D_{a}D_{b}K^{a}_{d}-D_{b}D_{a}K^{a}_{d}\right)\delta h^{bd}
\nonumber
\\
=&\Big[\frac{1}{2}\left(h_{dq}D_{a}D_{b}S^{ab}+D_{c}D^{c}S_{dq}\right)-D_{c}D_{d}S^{c}_{q}\Big]\delta h^{dq}
\nonumber
\\
&+4K^{p}_{d}\mathcal{R}_{pb}\delta h^{bd}-4K^{a}_{p}\mathcal{R}^{p}_{~dab}\delta h^{bd} \label{Expr1}
\end{align}
In the last step above, we have replaced the anti-commutator $[D_a,D_b]$ with the three-dimensional curvature using
\begin{align} \label{arbit}
D_{a}D_{b}A^{c}_{d}-D_{b}D_{a}A^{c}_{d} =\mathcal{R}^c_{~eab} A^e_d -\mathcal{R}^e_{~dab} A^c_e,
\end{align}
valid for any tensor $A^{a}_{b}$ that is orthogonal to the normal on both indices.
Proceeding to the last term in \ref{LL_m_2}, we have 
\begin{align}
\textrm{Expression 2}&=-\widetilde{S}_{a_{1}b_{1}\ (2,1)}^{c_{1}d_{1}}\mathcal{R}^{a_{1}}_{\phantom{a}e_{1}c_{1}d_{1}} \delta h^{e_1 b_1}=-\widetilde{S}^{cd}_{ab}\mathcal{R}^{a}_{~ecd}\delta h^{eb} \nn\\
&=-2\left[K\left(\delta ^{c}_{a}\delta ^{d}_{b}-\delta ^{c}_{b}\delta ^{d}_{a}\right)
-\left(K^{c}_{a}\delta ^{d}_{b}-\delta ^{c}_{b}K^{d}_{a}\right)
+\left(K^{c}_{b}\delta ^{d}_{a}-\delta ^{c}_{a}K^{d}_{b} \right)\right]\mathcal{R}^{a}_{~ecd}\delta h^{eb}
\nonumber
\\
&=-4K\mathcal{R}_{ab}\delta h^{ab}+4K^{c}_{a}\mathcal{R}^{a}_{~ecb}\delta h^{eb}+4K^{c}_{b}\mathcal{R}_{ec}\delta h^{eb}~. \label{Expr2}
\end{align}
Adding the two terms from \ref{Expr1} and \ref{Expr2}, we obtain
\begin{align}
\textrm{Expression 1+Expression 2}=&-2D_{f_{1}}D^{c_{1}}\widetilde{S}^{f_{1}}_{~e_{1}c_{1}b_{1}\ (2,1)}\delta h^{e_{1}b_{1}}-\widetilde{S}^{c_1 d_1}_{a_1 b_1\ (2,1)}\mathcal{R}^{a_1}_{~e_1 c_1 d_1}\delta h^{e_1 b_1}~~ \nn\\
=&~~\Big[\frac{1}{2}\left(h_{dq}D_{a}D_{b}S^{ab}+D_{c}D^{c}S_{dq}\right)-D_{c}D_{d}S^{c}_{q}\Big]\delta h^{dq}
\nonumber
\\
&+4K^{p}_{d}\mathcal{R}_{pb}\delta h^{bd}-4K^{a}_{p}\mathcal{R}^{p}_{~dab}\delta h^{bd}
\nonumber
\\
&-4K\mathcal{R}_{ab}\delta h^{ab}+4K^{c}_{a}\mathcal{R}^{a}_{~ecb}\delta h^{eb}+4K^{c}_{b}\mathcal{R}_{ec}\delta h^{eb}
\nonumber
\\
=&\Big[\frac{1}{2}\left(h_{dq}D_{a}D_{b}S^{ab}+D_{c}D^{c}S_{dq}\right)-D_{c}D_{d}S^{c}_{q}+8\left(K^{m}_{d}\mathcal{R}_{mq}-\frac{K}{2}\mathcal{R}_{dq}\right)\Big]\delta h^{dq}~.
\end{align}
Hence, the Dirichlet variation term in our general \LL also reduces to the \GB expression for $m=2$. 

\section{The Conjugate Momentum for the Gauss-Bonnet Case} \label{app:GB:Conj:Mom}

In this section, we shall compare the expression for conjugate momentum derived for \GB gravity with expressions existing in the literature (derived in \cite{Davis:2002gn,Gravanis:2002wy}, also see \cite{Deruelle:2003ck} where the expression is explicitly written down). From \ref{GB_BT_final_sqh}, the conjugate momentum for \GB gravity reads
\begin{align}
\sqh \Pi _{dq}^{(2)}=&\sqh\left\{D^{b}B_{qbd~(2)}+\epsilon A^{b}_{d~(2)}K_{bq}+8\left(K^{m}_{d}\mathcal{R}_{mq}-\frac{K}{2}\mathcal{R}_{dq}\right)\right.\nn\\
&+\frac{1}{2}\left(h_{dq}D_{a}D_{b}S^{ab}+D_{c}D^{c}S_{dq}\right)-D_{c}D_{d}S^{c}_{q}
\nonumber
\\
&\left.-\frac{1}{2}h_{dq}\left[8 \mathcal{R}^{a}_{b}K^{b}_{a}-4 \mathcal{R}K+\epsilon\left(\frac{4}{3}K^{3}+\frac{8}{3}K^{a}_{b}K^{b}_{c}K^{c}_{a}-4KK^{a}_{b}K^{b}_{a}\right) \right]\right\}\vbar_{\textrm{sym}[d\leftrightarrow q]}~, \label{Pi_GB}
\end{align}
where we have indicated that this object should be symmetrized in indices $d$ and $q$ since it is contracted with $\df h^{dq}$. For \GB gravity, using \ref{def_B} and \ref{P_Gauss_Bonnet},
\begin{align}
B_{qbd~(2)}&=2P^{ancp}_{(2)}h_{qa}h_{bn}n_{c}h_{dp}
\nonumber
\\
&=4R^{ancp}h_{qa}h_{bn}n_{c}h_{dp}+4G^{nc}h_{bn}n_{c}h_{qd}-4G^{ac}h_{qa}n_{c}h_{bd}
\nonumber
\\
&=4\left(D_{q}K_{bd}-D_{b}K_{qd}\right)+4\left(D_{b}K-D_{a}K^{a}_{b}\right)h_{dq}-4\left(D_{q}K-D_{a}K^{a}_{q}\right)h_{bd}~.
\end{align}
Thus,
\begin{align}\label{DB_term}
D^{b}B_{qbd~(2)}&=4D^{b}D_{q}K_{bd}-4D^{b}D_{b}K_{qd}-4h_{dq}D^{a}D^{b}K_{ab}+4D_{d}D^{a}K_{qa}-4D_{d}D_{q}K+4h_{dq}D^{b}D_{b}K~.
\end{align}
Next, we shall simplify the second line of \ref{Pi_GB} using $S^{ab}=8(K^{ab}-(1/2)Kh^{ab})$:
\begin{align}
-D_{c}D_{d}S^{c}_{q}+\frac{1}{2}\left(h_{dq}D_{a}D_{b}S^{ab}+D_{c}D^{c}S_{dq}\right)=&-8D_{b}D_{d}K^{b}_{q}+4D_{q}D_{d}K+4\left(D_{a}D_{b}K^{ab}\right)h_{dq}
\nonumber
\\
&+4D_{c}D^{c}K_{dq}-4\left(D_{c}D^{c}K\right)h_{dq} \label{DS}
\end{align}
Adding \ref{DB_term} and \ref{DS} and symmetrizing in $d$ and $q$, we obtain
\begin{align}
\left[D^{b}B_{qbd~(2)}-D_{c}D_{d}S^{c}_{q}+\frac{1}{2}\left(h_{dq}D_{a}D_{b}S^{ab}+D_{c}D^{c}S_{dq}\right)\right]\vbar_{\textrm{sym}[d\leftrightarrow q]}&=4[D_{d},D_{a}]K_{q}^{a}\vbar_{\textrm{sym}[d\leftrightarrow q]}
\nonumber
\\
&=\left[4\mathcal{R}^{a}_{~sda}K^{s}_{q}-4\mathcal{R}^{s}_{~qda}K^{a}_{s}\right]\vbar_{\textrm{sym}[d\leftrightarrow q]}\nn\\&=\left[-4\mathcal{R}_{db}K^{b}_{q}+4\mathcal{R}_{dbqs}K^{bs}\right]\vbar_{\textrm{sym}[d\leftrightarrow q]},
\end{align}
where we have made use of \ref{arbit}. Our next target of simplification will be the term with $A^{b}_{d~(2)}$ in \ref{Pi_GB}. Making use of \ref{Adm_GB},
\begin{align}
\epsilon A^{b}_{d~(2)}K_{bq}=4\mathcal{R}^{b}_{d}K_{bq}-2\mathcal{R}K_{dq}-4\epsilon \left[-K^{b}_{c}K^{c}_{d}K_{bq}+KK^{b}_{d}K_{bq}
+\frac{1}{2}K_{dq}\left(K_{ab}K^{ab}-K^{2}\right)\right],
\end{align}
so that
\begin{align}
\nl\epsilon A^{b}_{d~(2)}K_{bq}-\frac{1}{2}h_{dq}\left[8 \mathcal{R}^{a}_{b}K^{b}_{a}-4 \mathcal{R}K+\epsilon\left(\frac{4}{3}K^{3}+\frac{8}{3}K^{a}_{b}K^{b}_{c}K^{c}_{a}-4KK^{a}_{b}K^{b}_{a}\right) \right]
\nonumber
\\
=&4\mathcal{R}^{b}_{d}K_{bq}-2\mathcal{R}K_{dq}-4\epsilon \left[-K^{b}_{c}K^{c}_{d}K_{bq}+KK^{b}_{d}K_{bq}
+\frac{1}{2}K_{dq}\left(K_{ab}K^{ab}-K^{2}\right)\right]
\nonumber
\\
&+h_{dq}\left(-4\mathcal{R}_{ab}K^{ab}+2K\mathcal{R}\right)+2\epsilon h_{dq}\left(-\frac{1}{3}K^{3}-\frac{2}{3}K^{b}_{c}K^{c}_{d}K^{d}_{b}+KK_{ab}K^{ab} \right)~.\label{arbit1}
\end{align}
Following \cite{Davis:2002gn}, we define the symmetric quantity $J_{ab}$ 
\begin{align}
J_{ab}=-\frac{2}{3}K_{ac}K^{cd}K_{db}+\frac{2}{3}KK_{ac}K^{c}_{b}+\frac{1}{3}K_{ab}\left(K_{pq}K^{pq}-K^{2} \right),
\end{align}
and its trace
\begin{equation}
J=-\frac{2}{3}K_{ac}K^{cd}K_{da}+KK_{ac}K^{c}_{a}-\frac{1}{3}K^{3}~.
\end{equation}
In terms of these quantities, \ref{arbit1} can be written as
\begin{align}
\nl\epsilon A^{b}_{d~(2)}K_{bq}-\frac{1}{2}h_{dq}\left[8 \mathcal{R}^{a}_{b}K^{b}_{a}-4 \mathcal{R}K+\epsilon\left(\frac{4}{3}K^{3}+\frac{8}{3}K^{a}_{b}K^{b}_{c}K^{c}_{a}-4KK^{a}_{b}K^{b}_{a}\right) \right]
\nonumber
\\
=& 4\left[-\frac{3}{2}\epsilon J_{dq}+\frac{1}{2}\epsilon Jh_{dq} \right]
+4\left[\mathcal{R}^{b}_{d}K_{bq}-\frac{1}{2}\mathcal{R}K_{dq}-h_{dq}\left(\mathcal{R}_{ab}K^{ab}-\frac{1}{2}K\mathcal{R}\right) \right]~.
\end{align}
Thus, we obtain
\begin{align}
\Pi_{dq}=&\left\{-4\mathcal{R}_{db}K^{b}_{q}+4\mathcal{R}_{dbqs}K^{bs}+4\left[-\frac{3}{2}\epsilon J_{dq}+\frac{1}{2}\epsilon Jh_{dq} \right]\right.
\nonumber
\\
&\left.+4\left[\mathcal{R}^{b}_{d}K_{bq}-\frac{1}{2}\mathcal{R}K_{dq}-h_{dq}\left(\mathcal{R}_{ab}K^{ab}-\frac{1}{2}K\mathcal{R}\right)\right]+8\left(K^{m}_{d}\mathcal{R}_{mq}-\frac{1}{2}K\mathcal{R}_{dq}\right)\right\}\vbar_{\textrm{sym}[d\leftrightarrow q]}
\nonumber
\\
=&4\left[-\frac{3}{2}\epsilon J_{dq}+\frac{1}{2}\epsilon Jh_{dq} \right]
\nonumber
\\
&+4\Big[2\mathcal{R}^{b}_{d}K_{bq}-\frac{1}{2}\mathcal{R}K_{dq}+h_{dq}\left(-\mathcal{R}_{ab}K^{ab}+\frac{1}{2}K\mathcal{R}\right)+\mathcal{R}_{dbqs}K^{bs}-K\mathcal{R}_{dq} \Big]\vbar_{\textrm{sym}[d\leftrightarrow q]}
\nonumber
\\
=&4\left[-\frac{3}{2}\epsilon J_{dq}+\frac{1}{2}\epsilon Jh_{dq}+\frac{1}{2}\mathcal{P}_{dbqs\ (2)}K^{bs} \right],
\end{align}
where we have defined $\mathcal{P}_{dbqs\ (2)}$ as the boundary tensor analogous to $P^{ab}_{cd\ (2)}$ in \ref{P_Gauss_Bonnet}:
\begin{equation}\label{P_GB_boundary}
\mathcal{P}_{abcd\ (2)}=2\Big[\mathcal{R}_{abcd}+\mathcal{R}_{bc}h_{ad}-\mathcal{R}_{ac}h_{bd}-\frac{\mathcal{R}}{2}\left(h_{bc}h_{ad}-h_{ac}h_{bd}\right)+\mathcal{R}_{ad}h_{bc}-\mathcal{R}_{bd}h_{ac}\Big]~.
\end{equation}
In order to compare with the results derived in \cite{Davis:2002gn}, we shall put $\epsilon=1$ since the analysis there is done for a timelike boundary. Also, note that our definition of $P_{abcd\ (2)}$ in \ref{P_Gauss_Bonnet} is twice the expression defined as $P_{abcd}$ in \cite{Davis:2002gn}. Then, we have
\begin{align}
\Pi_{dq}=4\left[-\frac{3}{2} J_{dq}+\frac{1}{2} Jh_{dq}+\frac{1}{2}\mathcal{P}_{dbqs\ (2)}K^{bs} \right],
\end{align}
which matches with the expression obtained in \cite{Davis:2002gn} (written out explicitly in \cite{Deruelle:2003ck}) except for a minus sign.
This minus sign difference is again, as was the case with previous sections, due to difference in the convention for $K_{ab}$. Davis in \cite{Davis:2002gn} defines $K_{ab}=h^c_a\nabla_c n_b$, which differs from our definition (following \cite{gravitation,MTW}) by a minus sign. Taking this minus sign into account, our conjugate momentum matches the expressions provided in \cite{Davis:2002gn,Deruelle:2003ck}.

The conjugate momentum expression is also provided in \cite{Gravanis:2002wy}, where it is written in the form
\begin{equation}
\Pi_{dq}= -2\left(Q_{dq}-\frac{h_{dq}}{3}Q\right),
\end{equation}
with (the second term below needs to be symmetrized for symmetric $\Pi_{dq}$)
\begin{equation}
Q_{ab}= -2 K^{cd}\mathcal{R}_{acbd}- 4 \mathcal{R}_{ac}K^c_b+2 K \mathcal{R}_{ab}+\mathcal{R}K_{ab}+K_{ab}\left(K^{cd}K_{cd}-K^2\right)+2KK_{ac}K^c_b-2K^c_{a}K_{cd}K^d_b~.
\end{equation}
This expression is also easily seen to match our result.


\providecommand{\href}[2]{#2}\begingroup\raggedright\endgroup
\end{document}